\documentclass[letterpaper,twocolumn,10pt]{article}
\usepackage{usenix}

\usepackage{tikz}
\usepackage{amsmath}
\usepackage{booktabs}
\usepackage{alltt}
\usepackage{paralist} 

\usepackage{enumitem}
\usepackage[most]{tcolorbox}
\usepackage{amssymb}
\usepackage{pifont}
\usepackage{caption}
\usepackage{appendix}
\usepackage{multirow}
\usepackage{filecontents}
\usepackage{tabularx}
\newcolumntype{C}[1]{>{\centering\arraybackslash}p{#1}} 
\newcolumntype{L}[1]{>{\RaggedRight\arraybackslash}p{#1}} 
\newcolumntype{R}[1]{>{\RaggedLeft\arraybackslash}p{#1}} 
\usepackage{booktabs}
\usepackage{ragged2e} 

\newcommand{\paragraphb}[1]{\noindent{\bf #1} }

\newcommand{\paragraphbe}[1]{\vspace{0.03in} \noindent{\bf \em #1} }

\usepackage{array}
\usepackage{tcolorbox, color}
\tcbuselibrary{skins,listings,breakable}

\definecolor{todoamber}{RGB}{230,126,34} 

\begin{document}
\pagenumbering{gobble}

\date{}

\title{Network-Level Prompt and Trait Leakage in Local Research Agents}


\author{
  {\rm Hyejun Jeong}\quad
  {\rm Mohammadreza Teymoorianfard}\quad
  {\rm Abhinav Kumar}\quad \vspace{0.1cm} \\ \vspace{0.1cm}
  {\rm Amir Houmansadr}\quad
  {\rm Eugene Bagdasarian} \vspace{0.2cm} \\ \vspace{0.1cm}
  University of Massachusetts Amherst\\
  \texttt{\{hjeong,mteymoorianf,abhinavk,amir,eugene\}@cs.umass.edu}
}
\newcommand\blfootnote[1]{%
  \begingroup
  \renewcommand\thefootnote{}\footnote{#1}%
  \addtocounter{footnote}{-1}%
  \endgroup
}

\maketitle

\begin{abstract}
We show that Web and Research Agents (WRAs)---language-model-based systems that investigate complex topics on the Internet---are vulnerable to inference attacks by passive network observers. Deployment of WRAs \emph{locally} by organizations and individuals for privacy, legal, or financial purposes exposes them to DNS resolvers, malicious ISPs, VPNs, web proxies, and corporate or government firewalls.
However, unlike sporadic and scarce web browsing by humans, WRAs visit $70{-}140$ domains per each request with a distinct timing pattern creating unique privacy risks.

Specifically, we demonstrate a novel prompt and user trait leakage attack against WRAs that only leverages their network-level metadata (i.e., visited IP addresses and their timings). 
We start by building a new dataset of WRA traces based on real user search queries and queries generated by synthetic personas.
We define a behavioral metric (called OBELS) to comprehensively assess similarity between original and inferred prompts, showing that our attack recovers over 73\% of the functional and domain knowledge of user prompts.
Extending to a multi-session setting, we recover up to 19 of 32 latent traits with high accuracy.
Our attack remains effective under partial observability and noisy conditions.
Finally, we discuss mitigation strategies that constrain domain diversity or obfuscate traces, showing negligible utility impact while reducing attack effectiveness by an average of 29\%.
\end{abstract}

\blfootnote{Code, prompts available at \url{https://github.com/umass-aisec/wra}}

\begin{figure*}[ht]
    \centering
    \includegraphics[width=1.0\linewidth]{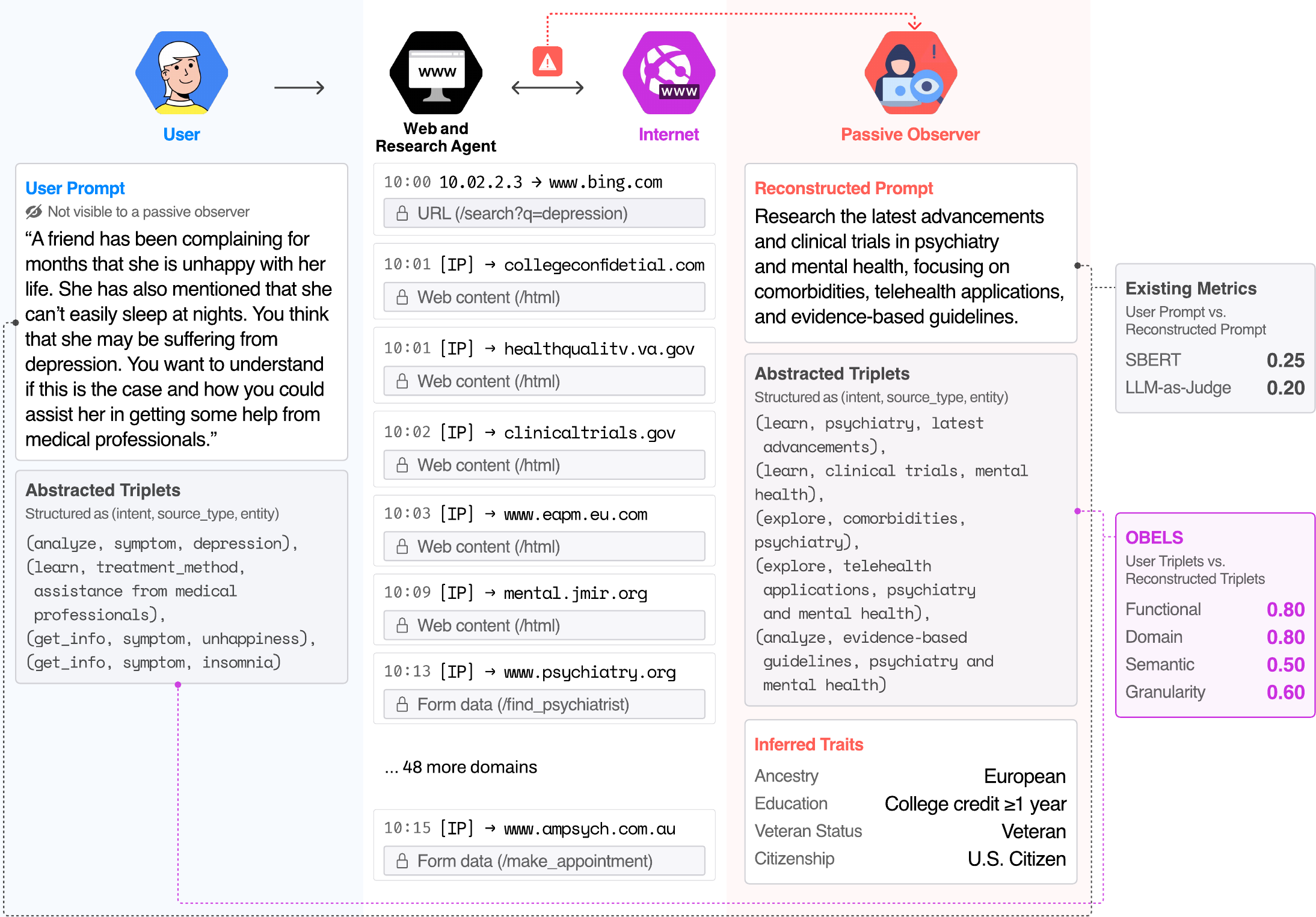}
    \caption{\textbf{Overview of the attack.}}
    \label{fig:threat_model}
\end{figure*}

\section{Introduction}

Web and Research Agents (WRAs) are reshaping how users interact with information. 
Unlike traditional assistants that answer isolated queries, these Large Language Model (LLM)-powered systems autonomously plan, browse, and synthesize knowledge across the web~\cite{huang2025deep}. 
Major AI providers are integrating such agents directly into their flagship assistants or browsers, including OpenAI \cite{openai2025introducingdeepresearch}, Google Gemini \cite{google2025geminideepresearch}, Grok \cite{xai2025grok3}, Mistral \cite{mistral2025lechat}, and Perplexity \cite{perplexity2025deepresearch}.
Open-source counterparts such as LangChain's Local Deep Researcher \cite{langchain2025localdeepresearcher}, GPT Researcher~\cite{elovic2023gptresearcher}, and Agent Laboratory~\cite{schmidgall2025agent} are also widely adopted in academic and developer communities. 
Their growing deployment in assistants, productivity platforms, and enterprise environments, particularly within privacy-sensitive sectors such as healthcare, law, and finance, raises urgent concerns about their security and privacy implications.
While on-device execution through locally-deployable models and TLS-encrypted access could suggest privacy guarantees, this perception is misleading.
Yet WRAs leave unavoidable, structured metadata traces that can reveal sensitive user intent, creating an underexplored leakage vector.

As autonomous agent systems, WRAs inherit vulnerabilities from both LLMs and traditional web browsing
~\cite{hui2024pleak,carlini2021extracting,siby2019encrypted,bird2020replication}.
Yet they also expose a distinct, underexplored vulnerability: behavioral traces determined by an LLM acting on a user's query.
Prior work has shown that user intent and goals can be inferred from search queries and browsing activity
~\cite{rose2004understanding,jones2007know}.
Even when content is encrypted, metadata such as domain names, access order, payload size, and timing can silently reveal what a user is trying to accomplish
~\cite{oliveira2023browsing,su2017anonymizing}.

Crucially, these behavioral traces are operationally distinguishable.
WRAs generate dense cascades of semantically related domain visits within short time windows \cite{perplexity2025deepresearch}.
Unlike cognitively paced, irregular, and opportunistic human browsing \cite{kellar2007field}, WRAs follow deterministic retrieval-and-synthesis loops with low variance in per-step timing.
As a result, a single WRA session often traverses 40–100 distinct domains, whereas human sessions typically rely on few sites \cite{wang2025ai}.
These structured, high-throughput access patterns make WRA traffic reliably separable from human activity and particularly amenable to profiling and inference attacks \cite{chiang2025web,mudryi2025hidden}.

We adopt a realistic threat model in which a passive adversary, (e.g., DNS resolver, ISP, enterprise firewall, or local network operator) observes only domain-level traffic metadata, without access to prompts, page contents, or model internals. 
Despite this limited visibility, we show that adversaries can reliably recover what users asked and who they are.  

\paragraphbe{Motivating Examples.}
Consider a user seeking reproductive health information: their research agent may visit clinics, support forums, and medical sites. 
Even without access to page contents, these traces alone can expose the agent’s internal reasoning process and deeply stigmatized intent. 
As WRAs become integrated into daily decision-making, their behaviors present a new leakage channel. 
In essence, privacy failures emerge not from what agents say, but from how they act. 

\paragraphbe{Our Work.}
We present privacy leakage attacks that exploit metadata traces of WRAs deployed locally. 
The attacks include \textbf{(1) Prompt Recovery}, which recovers a user’s original prompt from observed domain sequences, and \textbf{(2) Trait Inference}, which profiles latent attributes (e.g., gender, ideology) from the agent's multi-session browsing patterns.  
The adversary trains an inference model using task-specific strategies such as In-Context Learning (ICL) and fine-tuning to map browsing traces to inference targets (\autoref{tab:finetuning_compar}).
Our results show that the attacks remain effective even when 40\% of the traces are masked or fully obfuscated, revealing that encrypted content alone cannot guarantee privacy (\autoref{tab:prompt_rec_matadata}).
We observe that WRAs visit dozens of marginal or redundant domains that do not improve the final report but still expand the adversary’s visibility (\autoref{tab:prompt_rec_utility}).
To mitigate these risks, we propose and evaluate practical defenses that either \textbf{hide traces}, by injecting LLM-generated decoy prompts and virtual personas to confuse inference (Section~\ref{sec:defense-hiding}), or \textbf{block traces}, by redirecting tasks to multipurpose sources or LLM knowledge, avoiding uniquely identifying domains (Section~\ref{sec:defense_blocking}).

Our contributions are as follows: 
\begin{compactitem}
    \item \textbf{New attack surface:} We introduce metadata-based privacy attacks from behavioral traces produced by WRAs and propose a general two-stage pipeline for prompt recovery and trait inference.
    \item \textbf{Benchmark dataset and OBELS metric:} We construct and release a dataset linking domain traces to prompts and persona traits, and introduce OBELS, an ontology-aware metric for evaluating privacy leakage through behavioral traces.
    \item \textbf{Ablation and mitigation:} We conduct extensive experiments across agents, LLMs, and masking conditions, and evaluate mitigation that constrain domain diversity or obfuscate traces while negligibly affecting utility.
\end{compactitem}

\section{Background and Related Work}

\subsection{AI, Web, and Research Agents}

\paragraphbe{Standalone LLMs to AI Agents.} 
LLMs excel at language understanding but are limited as static knowledge systems \cite{achiam2023gpt, team2024gemini}. 
They cannot access real-time information, call external APIs, or execute multi-step plans, but produce responses based only on their pretrained knowledge. 
Early solutions such as RAG \cite{lewis2020retrieval, guu2020retrieval} and tool-augmented LLMs \cite{schick2023toolformer, shen2023hugginggpt, qin2023toolllm, chen2024advancing} improved grounding and tool use, but remain constrained to pre-indexed corpora and invoking tools only when explicitly instructed. 
AI agents address this by embedding LLMs into a perception--action loop \cite{yao2023react, autogpt, wang2023voyager}, where the model plans, executes, and adapts across multiple steps, enabling autonomous reasoning and dynamic interaction with external environments.

\paragraphbe{Web Agents.}
Generic agents can reason about which queries to run, but often lack robust capabilities for controlling browsers, parsing dynamic content, handling JavaScript-heavy pages, or managing session state across multiple sites.  
Web agents integrate LLM planning with browser control: issuing search queries, navigating between sources, extracting content, and deciding which actions to perform based on the evolving context \cite{chiang2025web}. 
Frameworks like AutoGen~\cite{wu2023autogen} and Browser-Use~\cite{browser_use2024} exemplify these capabilities, though evaluations often emphasize narrow navigation tasks.

\begin{figure}
    \centering
    \includegraphics[width=.8\linewidth]{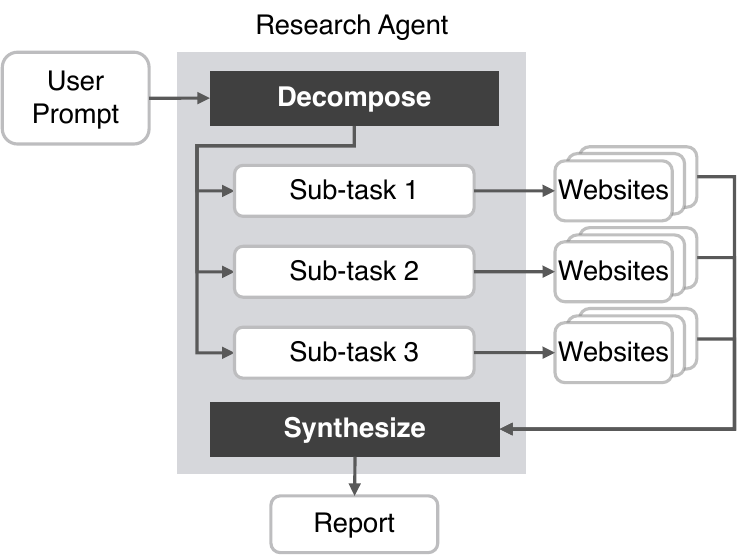}
    \caption{The workflow of Web and Research Agents.}
    \label{fig:research-agent-design}
\end{figure}

\paragraphbe{Research Web Agents (WRAs).}
WRAs operationalize deep evidence synthesis through a \emph{search--retrieve--extract--synthesize} loop across many sources (\autoref{fig:research-agent-design}). 
Given a user query, the agent generates sub-queries and issues them via APIs such as Tavily \cite{tavily} to minimize token overhead and retrieve relevant documents efficiently. 
Retrieved content is parsed into evidence, distilled into notes, and evaluated to decide whether further exploration is needed. 
This process repeats until sufficient coverage is achieved, at which point the agent composes a comprehensive, citation-rich output \cite{huang2025deep,schmidgall2503agentrxiv,zheng2025deepresearcher,perplexity2025deepresearch,chiang2025web}.  
Systems exemplifying this paradigm include GPT Researcher~\cite{elovic2023gptresearcher}, LangChain’s Local Deep Researcher~\cite{langchain2025localdeepresearcher}, OpenAI Deep Research~\cite{openai2025introducingdeepresearch}, Grok DeepSearch~\cite{xai2025grok3}, and Gemini Deep Research~\cite{google2025geminideepresearch}.

Because WRAs generate dense browsing traces over dozens of domains, they amplify metadata-leakage risks. 
This risk has been underscored when Grok and OpenAI misconfigurations exposed private prompts to public indexing~\cite{bbc2025example, stokel-walker2025google}. 
Locally deployed agents can be one of the mitigation strategies against such threats.
Running on personal machines or servers, they preserve confidentiality while offering lower latency, offline use, and lower cost. 
Examples include GPT-Researcher~\cite{elovic2023gptresearcher} with self-hosted backends such as Ollama~\cite{ollama2025} and AMD’s GAIA~\cite{amd_gaia_github} to operate fully on-device, giving users more control over both their data and their agent workflows.

\subsection{Privacy Attacks}

\paragraphbe{Threats in LLMs. }
LLMs introduce unique privacy risks due to their training data and the ways users interact with them.
One major class of attacks focuses on hidden-input inference, recovering sensitive inputs such as training data, system prompts, or user queries~\cite{carlini2021extracting, zhang2023effective, hui2024pleak, pasquini2024llmmap, liu2025evaluatingllm, carlini2024remote, weiss2024your}.
Other work has focused on recovering hidden system prompts used internally by providers \cite{zhang2023effective, hui2024pleak}, where adversaries probe model input and output patterns to elicit protected instructions.
Unlike these approaches, our setting assumes \textit{no direct access to the model} and relies instead on indirect behavioral signals.

Beyond direct extraction, studies show that semantic leakage occurs even when the attacker only observes generated outputs.
Pasquini et al.\cite{pasquini2024llmmap} and Liu et al.\cite{liu2025evaluatingllm} found that reconstructed prompts often preserve enough semantic content for intent inference, while Carlini et al.\cite{carlini2024remote} and Weiss et al.\cite{weiss2024your} demonstrated that response timing and token-length patterns can reveal coarse-grained information about prompt topics.
These attacks, however, assume access to model outputs or fine-grained response timings.
By contrast, we show that \textbf{passively observable network-level metadata alone}, without accessing queries or responses, can be sufficient to infer hidden prompts.

LLMs also introduce privacy risks through trait inference and user profiling.
Their outputs carry author-identifying signals~\cite{narayanan2012feasibility, schwartz2013personality, hovy2015demographic, azucar2018predicting}, allowing inference of attributes such as age, gender, and personality. 
Other risks include speculative decoding revealing fragments of private inputs~\cite{wei2024privacy} and long-context memory retaining user data across sessions for later retrieval~\cite{wang2025unveiling}.
Unlike these works, we assume a \textbf{strictly weaker threat model}: the adversary does \textbf{not} observe model inputs, outputs, or internal states, relying solely on external traffic traces (e.g., domain sequences, timing, payload sizes).

\paragraphbe{Threats in Web Privacy. }
Metadata leakage has long been recognized as a central privacy risk on the web.
Classical work on anonymous communications (e.g., MIX Nets~\cite{chaum1981untraceable} and Tor~\cite{dingledine2004tor}) demonstrated that encrypting content is insufficient when traffic patterns remain exposed.
Website Fingerprinting (WF) studies~\cite{wright2009traffic, cai2012touching, hayes2016k, juarez2016toward} showed that adversaries can infer visited pages based on encrypted flow patterns, while large-scale deanonymization attacks like the Netflix attack\cite{narayanan2008robust} revealed that sparse behavioral traces are often enough to re-identify individuals.
Subsequent work confirmed that traffic features, such as packet sizes, burst patterns, and inter-arrival timing, can fingerprint websites even under protections like Tor \cite{hintz2002fingerprinting, panchenko2011website}.
Parallel work in web tracking exposed persistent profiling vectors, including cookie-based tracking~\cite{lerner2016internet}, cookieless fingerprinting~\cite{nikiforakis2013cookieless, acar2014web}, and evercookies~\cite{ayenson2011flash}.

Recent work has shifted to domain-level metadata as a source of behavioral leakage.
Oliveira et al.\cite{oliveira2023browsing} showed that as few as four domains can uniquely re-identify most users in real-world datasets.
Crichton et al.\cite{crichton2025rethinking} found that domain sequence fingerprints remain stable across years, even after cookie resets.
Cross-context linkage is also possible: Naini et al.\cite{naini2015you} and Su et al.\cite{su2017anonymizing} demonstrated that anonymized browsing logs can be matched across platforms, breaking assumed protections. 
Whereas classical WF identifies \emph{which} site a user visits, our attack infers \emph{user's task or intent} from agent-driven traversals. 
The two are complementary: WF reveals destinations, while our method reconstructs prompts or traits from the resulting domain sequences.

While protocol-level defenses such as HTTPS, DNS-over-HTTPS (DoH), and DNS-over-TLS (DoT) aim to hide domain names too, they often fail to conceal key identifiers.
Hoang et al.\cite{hoang2020assessing} showed that TLS handshakes frequently leak domain information, while Sunahara et al.\cite{sunahara2025framework} pointed out widespread misconfigurations in encrypted DNS deployments, enabling persistent metadata leakage.
Similarly, Wang et al.~\cite{wang2025wechat} demonstrated that in-app telemetry (e.g., navigation events, click logs, and tracking APIs) leaks fine-grained behavioral patterns to platforms.
Unlike these works, we assume a weaker adversary: a passive network observer without privileged telemetry access, relying only on network-level domain sequences and packet metadata.

\paragraphbe{Agentic Privacy and Security Threats.}
AI agents amplify privacy risks by combining LLM reasoning with active browsing and external tool integration \cite{googlereport}.
Kim et al.~\cite{kim2024llms} showed that compromised agents can scrape PII from visited pages and reuse it in phishing, while InjecAgent~\cite{zhan2024injecagent} demonstrated that crafted payloads in webpages can trigger secret retrieval or unauthorized service access. 
Defenses such as AirGapAgent~\cite{bagdasarian2024airgapagent} and CHeaT~\cite{ayzenshteyncloak} aim to counter these threats through context isolation, behavior cloaking, and honey-token traps.

However, these prior studies assume an adversary with visibility into the agent’s queries, outputs, or injected prompts.
In contrast, our work identifies a \textbf{new and underexplored attack surface}: even without access to agent inputs, outputs, or model internals, \textbf{domain sequences and packet-level metadata alone} can be exploited to reconstruct hidden prompts and infer sensitive user traits.
This highlights a distinct dimension of privacy leakage introduced by web and research agents, where autonomous decision-making produces metadata-rich browsing patterns vulnerable to adversarial inference.
To our knowledge, this is the first study to identify and exploit metadata-driven leakage in WRAs.

\section{Overview: Prompt and Trait Leakage Attack} \label{sec:prompt-and-trait-leakage}
We study a \emph{passive network adversary} that monitors encrypted traffic between the AI agent (acting on behalf of the user) and the internet. 
Without access to page contents or search queries, the adversary can exploit metadata leakage to recover sensitive information about the user’s inputs and attributes.

\paragraphbe{Problem Definition.} 
WRAs execute user prompts by issuing queries, following links, and retrieving information, leaving behind browsing traces that encode sensitive signals. 
Even if the underlying page contents or prompts are hidden, the sequence and diversity of domains can reveal the intent behind a query, while recurring patterns across sessions expose persistent user traits. 
An adversary has clear incentives to exploit these inferences: prompt recovery compromises user intent privacy, while trait inference enables long-term profiling that can be monetized, surveilled, or used for manipulation.

\emph{Prompt Recovery.} 
With proxy dataset $D_{\text{PR}}=\{(WRA(p_j), p_j)\}$ consisting of pairs indexed by $j$, the adversary constructs an inference model $M_{\text{PR}}$.
For a new prompt $p$ with trace $t = WRA(p)$ (optionally filtered to $\tilde t$), the adversary recovers the prompt $\hat p$, aiming to maximize functional similarity:
\[
\hat p = M_{\text{PR}}(\tilde t), \quad
\max \; \mathrm{Sim}\big(M_{\text{PR}}(\tilde t),\, p\big).
\]

\emph{Trait Inference.} 
Let $\tau$ denote a user’s latent trait vector and $\{p_i\}_{i=1}^{N}$ the prompts across $N$ sessions, yielding $t^{1:N}=WRA(\{p_i\}_{i=1}^{N})$. 
From proxy dataset $D_{\text{TI}}=\{(WRA(\{p_i^{(j)}\}_{i=1}^{N}),\, \tau_j)\}$, the adversary instantiates an inference model $M_{\text{TI}}$.
On a new multi-session trace $t^{1:N}$ and infers trait $\hat\tau$, maximizing agreement:
\[
\hat\tau = M_{\text{TI}}(\tilde t^{\,1:N}), \quad
\max \; \mathrm{Score}\big(M_{\text{TI}}(\tilde t^{\,1:N}),\, \tau\big).
\]
where $\mathrm{Sim}$ and $\mathrm{Score}$ denote task-appropriate semantic similarity and trait-agreement metrics, respectively.

\subsection{Threat Model}
\paragraphbe{Adversary Goals.}
The adversary's objective is to infer sensitive user information from the browsing trace. 
We focus on two representative leakage goals that capture both short- and long-term risks.
The first is \textbf{prompt recovery}, where the adversary seeks to infer queries functionally equivalent to the user’s original input (e.g., detecting that a user asked about ``signs of depression and seeking professional mental health support''). 
The second is \textbf{trait inference}, where the adversary profiles latent attributes such as gender, religion, and political ideology by correlating patterns across multiple browsing sessions.  
Together, these illustrate how metadata can expose both immediate user intent and persistent personal characteristics.

\paragraphbe{Adversary Capabilities.}
The adversary observes domain-level metadata generated by the agent’s web activity, including domain names (via plaintext DNS or TLS handshake metadata such as SNI), the order of domain visits, and coarse packet-level features such as timing and payload sizes.
The adversary does \emph{not} observe full URLs (e.g., query strings), page contents, user prompts, or the agent’s intermediate queries.

In practice, visibility may be further reduced by Content Delivery Networks (CDNs), reverse proxies, or load balancers that aggregate multiple services under a small number of endpoints.
To capture such weaker settings, we explicitly evaluate partial-visibility scenarios in which only a fraction of domains (e.g., 80\%–10\%) is observable (\autoref{tab:prompt_rec_utility}).
These capabilities correspond to realistic on-path observers such as local ISPs, corporate firewalls, DNS resolvers, VPN providers, or shared network operators that collect encrypted metadata without decryption or manipulation.

\paragraphbe{Adversary Assumptions.}
We assume the user interacts with a local web or research agent that autonomously performs web browsing in response to a natural language prompt.
The adversary is aware of the agent’s general browsing capabilities, such as issuing web searches, clicking links, scrolling pages, or scraping structured data from web pages.
The adversary monitors the session between the agent and the Internet but does not alter traffic. 
We treat full-session metadata visibility as an \textbf{upper bound}: weaker adversaries such as DNS resolvers or enterprise firewalls may see only subsets of domains, yet our experiments show that WRA traffic remains highly distinguishable due to its structured, high-volume access patterns. 
We further assume that DNS and TLS handshake metadata remain visible (i.e., no DoH, DoT, or ECH) and that users do not employ VPNs, Tor, or other anonymizing proxies that would conceal domain sequences completely.

\subsection{Attack Scenarios} \label{sec:attack-scenarios}
In this section, we illustrate when a passive adversary would actually deploy our prompt and trait leakage attack. 
Locally deployed WRAs are especially relevant in settings where users avoid reliance on remote APIs, reduce recurring costs, maintain control over proprietary workflows, or process highly confidential information that cannot be shared externally. 
In such cases, browsing traffic is generated directly on the client machine, exposing metadata to on-path observers.

\paragraphbe{Personal WRAs.} An individual running a local WRA for personal tasks, such as health research, financial planning, or private study, may assume that confidentiality is preserved by avoiding cloud services. 
Yet even with local deployment, ISPs, enterprise firewalls, or other intermediaries can still observe traces and reconstruct prompts or infer private traits, enabling targeted advertising, discrimination, or phishing.

\paragraphbe{Academic WRAs.} In university, students and faculty use local WRAs for daily tasks, coursework, or research, leaking information about politics or unpublished work.
External observers like ISPs or competing research entities could exploit traces to reconstruct prompts or profile academic interests.

\paragraphbe{Industrial WRAs.} At the company level, employees use local WRAs for competitor analysis, strategic planning (e.g., product roadmaps, M\&A interests, financial planning), or handling confidential information (e.g., legal case files or patient records).
These traces can be mined for business intelligence by competitors, ISPs, cloud providers, or even internal IT monitors.
In startup companies, small teams could rely on local WRAs for cost savings or planning, but this may lead to leaks about funding strategies, partnership negotiations, or technical stack decisions.
ISPs or data brokers could harvest and resell the information or insights to competitors. 

\paragraphbe{State WRAs.} At the government or state level, state-operated firewalls or regulators monitoring domestic traffic could reconstruct politically sensitive prompts or identify activists and vulnerable communities for closer surveillance.

Overall, local deployment makes browsing activity directly observable in ways that remote, server-side agents do not. 
For adversaries, such visibility creates actionable opportunities to reconstruct immediate user prompts or build long-term trait profiles, depending on the target and setting.

\section{Attack Methodology} 
We present the end-to-end pipeline for our privacy attacks against AI agents. 
Both tasks---prompt recovery and trait inference---follow a shared two-stage process: 
(1) \textbf{offline model construction} on proxy datasets $D_{\text{PR}}$ or $D_{\text{TI}}$, yielding $M_{\text{PR}}$ or $M_{\text{TI}}$; and 
(2) \textbf{online attack execution} on real user traces $t$ or $t^{1:N}$ to produce $\hat p$ or $\hat\tau$.
The pipeline remains structurally consistent across tasks, though the input sources and inference targets differ.

\begin{figure}
    \centering
    \includegraphics[width=\linewidth]{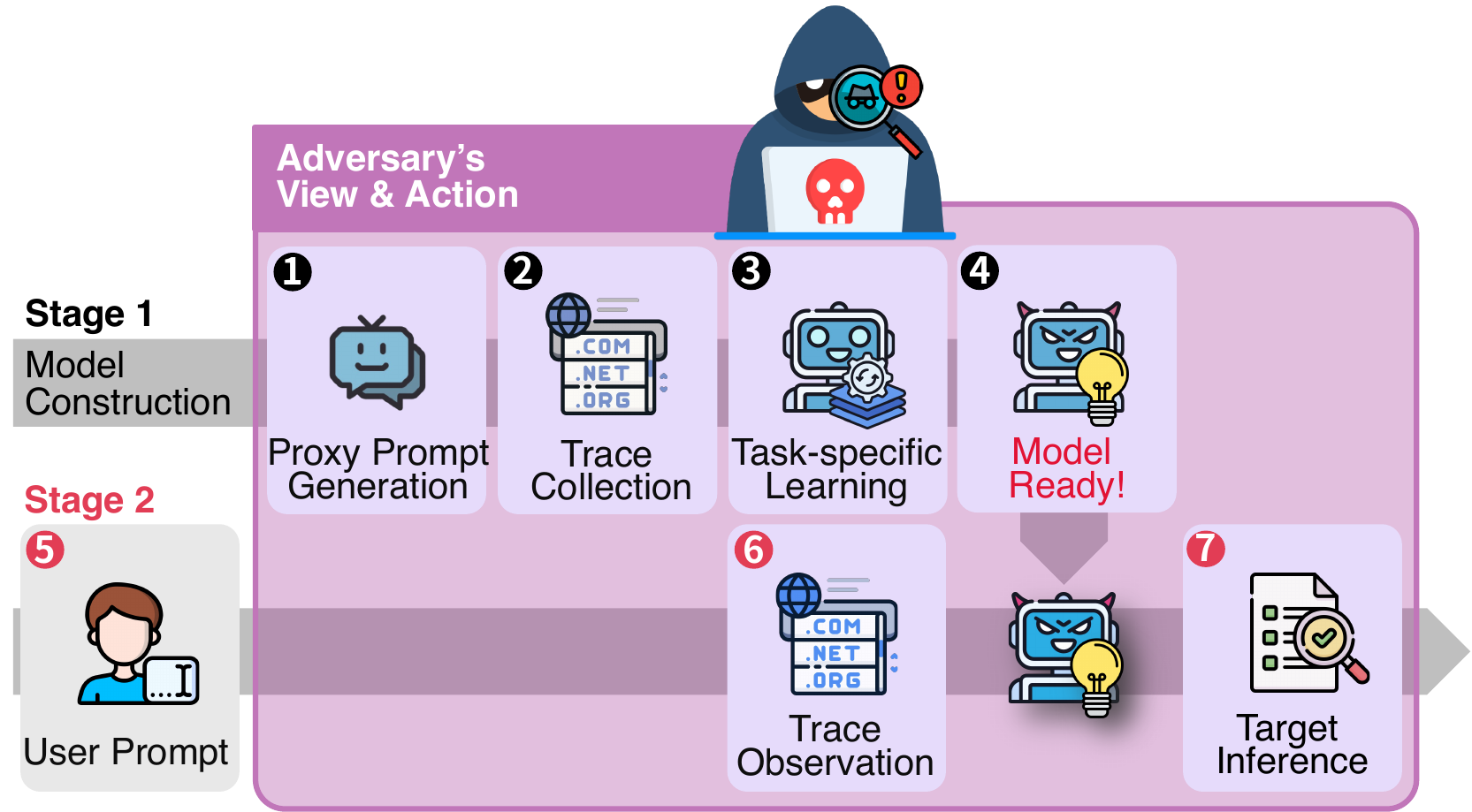}
    \caption{\textbf{Two-stage attack pipeline.}}
    \label{fig:attack_pipeline}
\end{figure}

\autoref{fig:attack_pipeline} illustrates our two-stage attack pipeline, separating the adversary’s model construction from the attack execution.  
A two-stage design is necessary because the adversary cannot directly obtain a large set of real user prompts paired with their browsing traces. 
Instead, the adversary first trains an inference model on proxy data that mimics real-world interactions, and then applies this trained model to actual users.  

\subsection{Proxy Prompt Generation}
To train the inference model in Stage 1, the adversary generates \textit{proxy prompts} that realistically simulate user behavior and produce informative traces $t$.
Effective proxy prompts must balance realism, browsing induction, and inferential utility. 
(a) Prompts resemble natural-language queries real users might issue (health, education, planning).
(b) They must reliably trigger multi-page browsing, rather than being satisfied by a single source (e.g., Wikipedia) or by the agent’s prior knowledge; to enforce this, we prepend instructions requiring at least five distinct page visits. 
(c) Prompts must be sufficiently detailed to produce distinguishable traces; underspecified inputs are rewritten into instruction-complete forms to avoid trivial queries that yield little signal. 
(d) Prompts and resulting traces remain semantically linked, so the trace encodes sufficient signal for inference. 

\paragraphbe{Prompt recovery.} 
We draw proxy prompts from public query datasets (FEDWEB13 \cite{fedweb13}, SESSION14 \cite{trec14-session-data}, DD16 \cite{trec16-domain-data}) to build $D_{\text{PR}}=\{(WRA(p_j),p_j)\}_j$. 
Some prompts are underspecified to reliably induce meaningful browsing, and agents differ in how they handle such inputs. 
GPT-Researcher \cite{elovic2023gptresearcher} internally refines vague prompts, but ChatGPT Deep Research UI asks clarifying questions, while Deep Research API, AutoGen, and Browser-Use execute inputs as-is. 
To align behaviors, Deep Research API prompts are rewritten into detailed variants (denoted -DR) to match the interactive interface.

\paragraphbe{Trait inference.} From persona profiles, we construct $D_{\text{TI}}=\{(WRA(\{p_i^{(j)}\}_{i=1}^{N}), \tau_j)\}_j$, embedding selected traits either explicitly or implicitly into prompts. 
This mirrors real user behavior, where personal context guides information seeking (e.g., health questions tied to age or history). 
A persistent adversary linking sessions over time (e.g., by IP) would observe recurring patterns tied to these traits ($\tau_j$). 
Executing such prompts through agents produces labeled traces that reveal stable user attributes, enabling training of trait inference models ($M_{\text{TI}}$). 
Dataset-specific sampling and generation details are described in Section~\ref{sec:datasets_selection}.

\subsection{Domain Trace Collection}
To characterize how different prompts induce distinct browsing behavior, we collect domain-level metadata from agents via Playwright or built-in system logs.
For each prompt $p$, this yields a trace $t=\langle(d_k,u_k,s_k)\rangle_{k=1}^{L}$, where $d_k$ denotes accessed domain, $u_k$ timestamp, and $s_k$ payload size.
The ordered trace captures whether an agent revisits the same domains or explores the set of sources during a single prompt execution.
Although TLS application data does not expose full URLs or domain names directly, domain attribution remains feasible through DNS queries, SNI fields in TLS handshakes, or correlated DNS/HTTPS timing, all of which are observable to ISPs or other on-path entities. 
This matches realistic adversarial capabilities today, though we note that CDN consolidation or future deployment of DoH/DoT/ECH could reduce visibility.

\paragraphbe{Noise Filtering.} \label{sec:noise-filtering}
A passive adversary observes all outbound requests, including primary domains, embedded resources, ads, and trackers. 
Because much of this traffic is incidental rather than agent-driven, an effective adversary would attempt to filter noise to isolate the domains most indicative of user intent. 
We apply domain-based noise filtering to remove well-known advertising and tracking services (e.g., \texttt{doubleclick.net}, \texttt{onetrust.com}, \texttt{taboola.com}). 
This represents the simplest plausible filtering approach and can be implemented using standard blocklists (e.g., EasyList \cite{easylist2025}). 
More advanced adversaries could employ refined techniques (e.g., traffic classification or resource analysis) to further separate incidental requests from meaningful domains, which we leave to future work.

\subsection{Inference Model Construction and Use}
The adversary builds an inference model from proxy datasets, then applies it to new traces simulating real-user activity.  

For \textbf{prompt recovery}, the task is to reconstruct a natural-language prompt that captures the functional intent of the user’s original input.
The adversary constructs an inference model ($M_{\text{PR}}$) from proxy (trace, prompt) pairs in the proxy dataset ($D_{\text{PR}}$), using either ICL, where a small set of examples is included in the inference prompt without parameter updates, or fine-tuning, where the same proxy pairs are used to train a direct mapping from traces to prompts (\autoref{tab:finetuning_compar}).
Given a user prompt $p$ producing trace $t=WRA(p)$ (optionally filtered to $\tilde t$ per Section~\ref{sec:noise-filtering}), the resulting model $M_{\text{PR}}$ is then applied to new traces to reconstruct real user prompts $\hat p = M_{\text{PR}}(\tilde t)$.

For \textbf{trait inference}, the goal is to infer latent user attributes ($\hat\tau$) from browsing traces ($\tilde t^{\,1:N}$). 
Here, the proxy dataset ($D_{\text{TI}}$) is generated, consisting of proxy (trace, traits) pairs for each annotated persona profile, and the adversary adopts ICL to construct $M_{\text{TI}}$, since it outperformed fine-tuning in the prompt recovery task. 
For a new user with sessions $\{p_i\}_{i=1}^{N}$ and trace $t^{1:N}=WRA(\{p_i\}_{i=1}^{N})$, the adversary output $\hat\tau = M_{\text{TI}}(\tilde t^{\,1:N})$.
This enables an adversary to recover persistent attributes such as demographics, ideology, or lifestyle 
from observed browsing activity.

\section{Ontology-aware Behavioral Leakage Score}
We introduce \textit{OBELS (Ontology-aware Behavioral Leakage Scores)}, a multi-metric scoring scheme that decomposes leakage into dimensions of preserved functionality, domain targets, and semantic entities, providing a structured alignment with attacker goals and real-world agent behavior.

\subsection{Limitations of Existing Metrics}
In AI agent contexts, leakage risk stems from \textit{functional equivalence}: prompt recovery is risky if it leads the agent to similar sites, services, or goals, despite differences in wording.
Existing metrics fall short because they assess what prompts \emph{say}, not what they \emph{do}.

\textbf{SBERT score}, computed as cosine similarity between sentence embeddings~\cite{reimers2019sentence}, captures high-level semantic alignment but suffers from paraphrastic inflation.
It misleadingly assigns high scores to verbose, generic, or loosely related prompts \cite{steck2024cosine,you2025semantics}, overlooking task-specific behavior or intent.

\textbf{LLM-as-Judge} methods leverage LLMs to rate similarity through natural language instructions. 
While more tolerant to paraphrase and flexible, they are prompt-sensitive and often lack interpretability \cite{razavi2025benchmarking,lee2024one}, producing scalar scores without indicating which semantic elements were preserved or lost.

These shortcomings are evident in practice. 
For example, rephrasing ``Compare Italy and Spain’s digital nomad visas'' as ``How do work visas differ across southern Europe?'' yields low cosine similarity, yet the agent navigates to similar government portals and retrieves overlapping content. 
From a privacy standpoint, this is a successful reconstruction, yet traditional metrics fail to capture it.

\subsection{Semantic Triplet Scoring with OBELS}
To address these gaps, we propose \textbf{OBELS}, a set of four complementary scores designed to capture \textit{behavioral} alignment. 
Rather than collapsing leakage into a single scalar, OBELS decomposes alignment into distinct dimensions so that failures in one area (e.g., entity granularity) are not masked by strong similarity elsewhere. 
This decomposition yields interpretable diagnostics and aligns more directly with attacker goals and real-world agent behavior.
It does so by abstracting prompts into structured triplets:
\begin{center}
    \texttt{(intent, source\_type, entity)}
\end{center}

Here, \texttt{intent} denotes the high-level goal (e.g., learn, explore, analyze, compare, summarize, plan, decide, book, watch, read, evaluate); \texttt{source\_type} identifies the class of information or service sought (e.g., travel, symptom, policy\_area, event, visa\_process, treatment\_method, academic\_field, cuisine, etc); and \texttt{entity} specifies the core topic (e.g., Rome, Italy, depression, face transplant, immigration, PhD, Business, Swahili dish, etc).
Removing any triplet component collapses distinct browsing behaviors, whereas variations in tone or sentiment do not meaningfully alter domain traversal.

We evaluate each reconstruction across four dimensions:
\begin{itemize}[itemsep=0pt, topsep=0pt]
    \item \textbf{Functional Equivalence ($E_{func}$):}
    Do the two prompts express the same high-level intent or task (e.g., finding a product, booking a service)? 
    
    \item \textbf{Domain-Type Equivalence ($E_{domain}$):}  
    Do they engage with the same category of services or information sources (e.g., travel agencies, health databases)? 
    
    \item \textbf{Semantic Equivalence ($E_{semantic}$):}  
    Are entities semantically aligned (e.g., ``side effects of Prozac'' vs. ``adverse reactions to fluoxetine'')? 
    
    \item \textbf{Entity Granularity Tolerance ($T_{entity}$):}  
    Are differences in specificity (e.g., ``Italy'' vs. ``Rome and Venice'') minor enough to preserve meaning? 
\end{itemize}

Each dimension is scored from 0.0 (unrelated) to 1.0 (fully equivalent) using a standardized evaluation template applied to GPT-4, which compares triplet sets holistically and assigns fine-grained similarity scores.

Different applications may weight these components differently. 
For instance, entity granularity and semantic precision dominate in high-risk healthcare scenarios; domain-type alignment matters most in regulatory tasks; and intent alignment is sufficient for lower-risk consumer scenarios such as travel planning.
Collapsing these dimensions into a single score would obscure such task-sensitive variation and misrepresent leakage risk.

\begin{table*}[t]
\centering
\caption{\textbf{Traces of Web and Research Agents collected from different base datasets.}}
\renewcommand{\arraystretch}{0.8}
\small
\begin{tabular}{@{}lllclcclll@{}}
\toprule
\textbf{Base Dataset} & \textbf{Target} & \textbf{\# Total} & \textbf{\# Test} & \textbf{Agent} & \textbf{Avg. \# Dom} & \textbf{Avg. \# URLs} & \textbf{Prompts $\times$ Sessions} \\
\midrule
FEDWEB13 \cite{trec13-fedweb-topics} & Prompt & 50 prompts & 20 & GPT-Researcher & 77.5 & 143.0 & 1 prompt $\times$ 1 session \\ \midrule
SESSION14 \cite{trec14-session-data} & Prompt & 60 prompts & 20 & GPT-Researcher & 69.9 & 145.9 & 1 prompt $\times$ 1 session \\ \midrule
DD16 \cite{trec16-domain-data}       & Prompt & 53 prompts & 20 & GPT-Researcher & 64.8 & 155.8 & 1 prompt $\times$ 1 session \\ \midrule
PERSONA \cite{castricato2024persona} & Trait & 50 personas & 3  & AutoGen  & 144.1 & 173.7 & 3–5 prompts $\times$ 7 sessions \\
& &   (of 997) & &  & & & = 21–35 queries \\
\bottomrule
\end{tabular}
\label{tab:dataset_stats}
\end{table*}

\section{Experimental Setup} 
We evaluate our metadata-based prompt and trait leakage attacks on held-out test prompts. 
The objective is to assess how well an adversary can reconstruct user prompts that induce the same downstream behavior as the originals, or infer sensitive latent traits from domain traces, thereby capturing both immediate intent leakage and longer-term profiling risks.

\subsection{Datasets} \label{sec:datasets_selection}

We target two scenarios: prompt and trait leakage.
\autoref{tab:dataset_stats} summarizes all datasets; full details are in Appendix \ref{app:dataset_details}.

\paragraphbe{Prompt Recovery (TREC).} 
We use three TREC datasets: FedWeb 2013 (50 topics) \cite{trec13-fedweb-topics}, Session 2014 (60) \cite{trec14-session-data}, and Dynamic Domain 2016 (53) \cite{trec16-domain-data}, referred to as FEDWEB13, SESSION14, and DD16. 
For FEDWEB13 and DD16, we concatenate the \texttt{<description>} and \texttt{<narrative>} fields to match SESSION14's instruction-style and prompt specificity. 
To align with OpenAI's Deep Research API \cite{openai2025introducingdeepresearch}, we also generate \textbf{-DR variants} by rewriting each prompt with GPT-4.1, while scoring is always against the original prompts. 
For each dataset, 20 prompts are reserved for testing, with the remainder used for training and ICL examples.

TREC query text may appear in LLM pre-training corpora, but the domain-level traces we evaluate on are newly collected and do not exist in any pre-training dataset. 
The attack exploits these fresh behavioral traces, not memorized prompt content.

\paragraphbe{Trait Inference (PERSONA).}
We sample 50 profiles from PERSONA\_subset \cite{castricato2024persona} (997 total), each annotated with 32 traits. 
For each, we select 5 traits and use GPT-4o to generate 3–5 prompts across 7 sessions, simulating a week of browsing.
Three personas are held out for ICL, and after filtering to ages 18–70, 35 remain for testing. 

\subsection{WRAs and Backbone LLMs} \label{sec:trace_collection}
We evaluate domain-trace leakage in settings where a passive adversary can observe network-level metadata (IPs, domains, timing, payload size). 
To capture the traces, we use three open-source WRAs and one proprietary API agent (details in Appendix~\ref{app:trace_collection}):

\noindent\textbf{GPT Researcher \cite{elovic2023gptresearcher}.}
An open-source autonomous research agent for deep multi-hop investigation. 
We run it in \texttt{deep} mode, which produces structured research traces. 
GPT Researcher strategically uses multiple LLM backbones, and we evaluate both GPT-family and open-source backbones for the prompt recovery task.

\noindent\textbf{Browser-Use \cite{browser_use2024}.}
An open-source web agent that automates browsing through a local browser. 
To ensure stable traces, we configure it to start from Bing and require at least five distinct page visits before summarization. 
Browser-Use is used for the prompt recovery task with GPT-4o as the backbone.

\noindent\textbf{AutoGen \cite{wu2023autogen}.}  
An open-source web agent system we used for both prompt recovery (GPT-4o) and trait inference (Gemini 2.0 Flash), instrumented to log domains and metadata. 
Unlike the other agents, it produces the real packet-level traces, noisier with ads and trackers.

\noindent\textbf{OpenAI Deep Research API.}
A proprietary research agent that serves as a commercial baseline. 
To fully exercise its capabilities, we provide the \textbf{-DR} prompt variant (e.g., SESSION14-DR), designed to approximate the behavior of accessing Deep Research through the ChatGPT interface.  
While browsing is remote and not locally visible, it serves as a benchmark for high-capability proprietary systems.

Each open-source agent is paired with both open and proprietary LLM backbones to reflect realistic local deployments. 
OpenAI Deep Research API serves as a proprietary baseline. 
Larger local LLMs (e.g., LLaMA 3.1 70B, DeepSeek-R1) are excluded because they require substantial resources and often bypass web search by relying on internal knowledge.

\subsection{Evaluation Metrics} \label{sec:metrics}

\paragraphbe{Metrics for Prompt Recovery.}
We evaluate prompt recovery performance using three metrics:  
(1) \textbf{SBERT}, computed as cosine similarity over mean-pooled Sentence-BERT embeddings~\cite{reimers2019sentence};  
(2) \textbf{LLM-as-Judge}
, where an LLM rates prompt similarity based on natural language instructions; and  
(3) \textbf{OBELS}, our proposed behavior-aware scoring scheme to capture functional equivalence (i.e., whether the reconstructed prompt would induce similar agent behavior). \\

\paragraphbe{Metrics for Trait Inference.}
For trait inference, we adopt a type-aware scoring strategy tailored to each trait’s structure: 
\begin{compactitem}
    \item \textit{Numeric traits} are evaluated using a normalized absolute difference: smaller relative errors yield higher scores, linearly decreasing toward zero. 
    \item \textit{Ordinal traits} are mapped to integer levels and scored by normalized distance on a 5-point scale. Closer ratings yield higher similarity scores.
    \item \textit{Categorical traits} use exact match for single-token values; multi-word categories are scored using SBERT to account for semantic equivalence.
    \item \textit{Free-text traits} are assessed using SBERT, which computes semantic overlap via contextual token embeddings.
\end{compactitem}

This trait-type–aware evaluation avoids penalizing structured attributes with embedding-based metrics and ensures robust semantic assessment for open-ended descriptions. 
In contrast to prompt recovery, trait inference does not require alignment in downstream behavior; semantic similarity alone suffices to assess privacy leakage.

\section{Attack Results}

\subsection{Prompt Recovery}
Attack success is measured not by textual similarity alone, but by semantic and downstream behavior similarity, whether reconstructed prompts preserve the functional intent behind the original (see Section \ref{sec:metrics}).
The results quantify how much information about a user's input is revealed through observable browsing activity, and how the leakage rate varies across learning strategies, datasets, agents, and agents' LLM backbones. 
GPT-4o is used as the inference LLM unless stated otherwise. 
Appendix~\ref{app:inference_config} details the ICL setup, and additional results are reported in \autoref{app:additional-prompt_recon}.

\begin{table}[tp]
\caption{\textbf{5-shot ICL vs. supervised fine-tuning across inference LLMs.} ICL outperforms fine-tuning on all metrics, making it the stronger approach for prompt leakage attack.}
\label{tab:finetuning_compar}
\centering
\small
\renewcommand{\arraystretch}{0.9}
\setlength\tabcolsep{1pt}
\begin{tabular}{>{\raggedright\arraybackslash}m{2.7cm}
                    >{\centering\arraybackslash}m{1cm}
                    >{\centering\arraybackslash}m{1cm}
                    *{4}{>{\centering\arraybackslash}m{.8cm}}}
\toprule
\multirow{2}{*}[-0.8ex]{\parbox{2.7cm}{\centering \textbf{Models (Method)}}} &
\multirow{2}{*}[-0.8ex]{\parbox{1cm}{\centering \textbf{SBERT}}} &
\multirow{2}{*}[-0.8ex]{\parbox{1cm}{\centering \textbf{LLM-Judge}}} &
\multicolumn{4}{c}{\textbf{OBELS}} \\
\cmidrule{4-7}
 & & & \textbf{$E_{\text{func}}$} & \textbf{$E_{\text{dom}}$} & \textbf{$E_{\text{sem}}$} & \textbf{$T_{\text{ent}}$} \\ \midrule
GPT-4o (ICL)  & \textbf{0.492} & \textbf{0.415} & \textbf{0.770} & \textbf{0.735} & 0.520 & 0.640 \\  
GPT-4o (FT)    & 0.469 & 0.340 & 0.675 & 0.690 & 0.485 & 0.615 \\
GPT-4.1 nano (ICL) & 0.430 & 0.330 & 0.735 & 0.700 & 0.500 & 0.665 \\
GPT-4.1 nano (FT)    & 0.369 & 0.210 & 0.610 & 0.605 & 0.440 & 0.510 \\
Llama 3.1-8B (ICL) & 0.466 & 0.310 & 0.695 & 0.675 & 0.475 & 0.630 \\
Llama 3.1-8B (FT) & 0.416 & 0.230 & 0.705 & 0.710 & 0.470 & 0.525 \\
Qwen 2.5-32B (ICL) & 0.489 & 0.400 & 0.770 & 0.72 & \textbf{0.525} & \textbf{0.670} \\
Qwen 2.5-32B (FT)  & 0.404 & 0.225 & 0.595 & 0.640 & 0.410 & 0.530 \\ 
Gemma 3-4B (ICL)  & 0.429 & 0.280 & 0.675 & 0.670 & 0.475 & 0.635 \\
Gemma 3-4B (FT)  & 0.335 & 0.175 & 0.510  & 0.485 & 0.355 & 0.420 \\
\bottomrule
\end{tabular}
\end{table}

\paragraphbe{WRA vs. Human Attribution Analysis.}
To assess whether WRA traffic is distinguishable from human browsing, we clustered agent traces together with human sessions from SESSION14 using only timing, burstiness, and domain-transition features. 
WRAs consistently remain highly separable with 0.86--1.0 cluster purity across topics and 97.9\% global classification accuracy.
WRAs thus form a distinct, easily identifiable behavioral mode, exposing a richer and more attributable metadata footprint than humans.


\paragraphbe{ICL vs. Fine-Tuning Across Models.}
To motivate our choice of ICL over fine-tuning as the primary task-specific approach, we compare their performance across closed-source models (GPT-4o and GPT-4.1 nano) and open-source models (Llama 3.1 8B, Gemma 3 4B, and Qwen 2.5 32B) (\autoref{tab:finetuning_compar}).
We create our final training set by combining the training samples from the prompt recovery datasets discussed in Section~\ref{sec:datasets_selection}.  
For closed-source models, fine-tuning is performed through OpenAI’s fine-tuning API~\cite{openaiFinetuningAvailable} for 3 epochs with batch size 1.
For open-source models, we apply LoRA~\cite{hu2022lora} with rank 16 and $\alpha=12$.
Results show that ICL generally outperforms fine-tuning across nearly all metrics, though models like Llama occasionally favor fine-tuning for functional and domain equivalence. We attribute this to limited sample diversity, which likely causes overfitting in fine-tuned models.

\begin{table}[tp]
\caption{\textbf{Prompt recovery performance depending on different numbers of ICL examples.} 5-shot offers the best overall balance across evaluation metrics.}
\label{tab:prompt_rec_ICL_shots}
\centering
\small
\renewcommand{\arraystretch}{0.9}
\setlength\tabcolsep{3pt}
\begin{tabular}{>{\raggedright\arraybackslash}m{1.5cm}
                    >{\centering\arraybackslash}m{1cm}
                    >{\centering\arraybackslash}m{1.4cm}
                    *{4}{>{\centering\arraybackslash}m{.8cm}}}
\toprule
\multirow{2}{*}[-0.8ex]{\parbox{1.5cm}{\centering \textbf{No.\ of Examples}}} &
\multirow{2}{*}[-0.8ex]{\parbox{1cm}{\centering \textbf{SBERT}}} &
\multirow{2}{*}[-0.8ex]{\parbox{1.4cm}{\centering \textbf{LLM-Judge}}} &
\multicolumn{4}{c}{\textbf{OBELS}} \\
\cmidrule{4-7}
 & & & \textbf{$E_{\text{func}}$} & \textbf{$E_{\text{dom}}$} & \textbf{$E_{\text{sem}}$} & \textbf{$T_{\text{ent}}$} \\ \midrule
0-shot      & 0.395 & 0.295 & 0.665 & 0.630 & 0.465 & 0.615 \\
5-shot   & \textbf{0.492} & \textbf{0.415} & \textbf{0.770} & \textbf{0.735} & 0.520 & 0.640 \\
8-shot    & 0.490 & 0.400 & 0.760 & 0.730 & \textbf{0.540} & 0.610 \\
12-shot    & 0.504 & 0.405 & 0.750 & 0.710 & 0.530 & 0.640 \\
15-shot    & 0.487 & 0.385 & 0.770 & 0.685 & 0.525 & \textbf{0.645} \\
\bottomrule
\end{tabular}
\end{table}

\paragraphbe{Impact of Example Count.} \autoref{tab:prompt_rec_ICL_shots} compares performance across different numbers of in-context examples. 5-shot ICL provides a balanced trade-off between functional equivalence, domain alignment, and semantic similarity, while avoiding the performance drop observed at smaller or greater counts.

\begin{table}[t]
\centering
\small
\renewcommand{\arraystretch}{0.9}
\setlength\tabcolsep{1.8pt}
\caption{\textbf{Prompt recovery performance on SESSION14 across WFAs.} All agents are evaluated on 20 prompts with 5-shot ICL, except Deep Research$^*$ (5 prompts, 3-shot). GPT-Researcher traces leak the most, while Deep Research shows comparable leakage despite fewer shots.}
\label{tab:agent_compar}
\begin{tabular}{>{\centering\arraybackslash}m{2.25cm}
                    >{\centering\arraybackslash}m{1cm}
                    >{\centering\arraybackslash}m{1.1cm}
                    *{4}{>{\centering\arraybackslash}m{.8cm}}}
\toprule
\multirow{2}{*}[-0.8ex]{\parbox{2.25cm}{\centering \textbf{Agent}}} &
\multirow{2}{*}[-0.8ex]{\parbox{1cm}{\centering \textbf{SBERT}}} &
\multirow{2}{*}[-0.8ex]{\parbox{1.1cm}{\centering \textbf{LLM-Judge}}} &
\multicolumn{4}{c}{\textbf{OBELS}} \\
\cmidrule{4-7}
 & & & \textbf{$E_{\text{func}}$} & \textbf{$E_{\text{dom}}$} & \textbf{$E_{\text{sem}}$} & \textbf{$T_{\text{ent}}$} \\ \midrule
AutoGen & 0.356 & 0.225 & 0.565 & 0.625 & 0.400 & 0.495\\
Browser-Use & 0.368 & 0.240 & 0.540 & 0.590 & 0.390 & 0.485\\
GPT-Researcher & 0.492 & \textbf{0.415} & \textbf{0.770} & \textbf{0.735} & \textbf{0.520} & \textbf{0.640} \\
Deep Research$^*$ & \textbf{0.574} & 0.360 & 0.680 & 0.720 & 0.480 & 0.640 \\
\bottomrule
\end{tabular}
\end{table}

\paragraphbe{Impact of Agents.} 
As described in Sections~\ref{sec:datasets_selection} and~\ref{sec:trace_collection}, we collect traces from GPT-Researcher, AutoGen, and Browser-Use (20 prompts, 5-shot ICL) and from Deep Research (5 prompts, 3-shot) using SESSION14 prompts. 
\autoref{tab:agent_compar} shows that GPT-Researcher traces leak the most, consistent with its broader domain coverage and richer source diversity. 
Deep Research exhibits comparable leakage despite fewer prompts and shots, while AutoGen and Browser-Use generate learner traces that result in weaker leakage. 
These differences highlight how agent design and exploration strategy directly shape the degree of leakage.

\begin{table}[t]
\centering
\setlength\tabcolsep{1pt}
\small
\renewcommand{\arraystretch}{0.9}
\caption{\textbf{Prompt recovery on trace by GPT-Researcher with different backbones across datasets.} Local backbones leak less than GPT-4 traces but remain informative.}
\setlength\tabcolsep{1.4pt}
\label{tab:prompt_rec_dataset_compar}
\begin{tabular}{>{\centering\arraybackslash}m{1cm}
                    >{\centering\arraybackslash}m{1.5cm}
                    >{\centering\arraybackslash}m{1cm}
                    >{\centering\arraybackslash}m{1cm}
                    *{4}{>{\centering\arraybackslash}m{.8cm}}}
\toprule
\multirow{2}{*}[-0.8ex]{\parbox{1cm}{\centering \textbf{Model}}} &
\multirow{2}{*}[-0.8ex]{\parbox{1.5cm}{\centering \textbf{Dataset}}} &
\multirow{2}{*}[-0.8ex]{\parbox{1cm}{\centering \textbf{SBERT}}} &
\multirow{2}{*}[-0.8ex]{\parbox{1cm}{\centering \textbf{LLM-Judge}}} &
\multicolumn{4}{c}{\textbf{OBELS}} \\
\cmidrule{5-8}
 & & & & \textbf{$E_{\text{func}}$} & \textbf{$E_{\text{dom}}$} & \textbf{$E_{\text{sem}}$} & \textbf{$T_{\text{ent}}$} \\ \midrule
GPT4 & SESSION & 0.492 & 0.415 & 0.770 & 0.735 & 0.520 & 0.640 \\
Local & SESSION & 0.512 & 0.355 & 0.735 & 0.745 & 0.520 & 0.630 \\
GPT4 & FEDWEB & 0.505 & 0.305 & 0.755 & 0.735 & 0.540 & 0.695 \\
Local & FEDWEB & 0.493 & 0.310 & 0.750 & 0.745 & 0.515 & 0.645 \\
GPT4 & DD & 0.453 & 0.190  & 0.585 & 0.615 & 0.430 & 0.545 \\
Local & DD & 0.445 & 0.180 & 0.550 & 0.635 & 0.430 & 0.515 \\
\bottomrule
\end{tabular}
\end{table}

\paragraphbe{Impact of Agent Backbone: Proprietary vs. Open-Source.}
Using GPT-Researcher, we evaluated traces across datasets with either GPT-4 or local open-source backbones (DeepSeek-V3, Mistral family). As shown in \autoref{tab:prompt_rec_dataset_compar}, GPT-4 traces yield higher recovery scores, reflecting stronger reasoning capacity and broader exploration. Local models leak less overall, but still provide exploitable cues. For the remainder of our ICL experiments, we use GPT-Researcher with a GPT-4 backbone on SESSION14, as this setup offers both representativeness and clearer leakage signals for analysis.

\begin{table}[ht]
\caption{\textbf{Prompt recovery with different ICL inference models.} Proprietary models yield the highest scores.}
\label{tab:prompt_rec_models}
\centering
\small
\renewcommand{\arraystretch}{0.9}
\setlength\tabcolsep{1.8pt}
\begin{tabular}{>{\raggedright\arraybackslash}m{2.4cm}
                    >{\centering\arraybackslash}m{1cm}
                    >{\centering\arraybackslash}m{1cm}
                    *{4}{>{\centering\arraybackslash}m{.8cm}}}
\toprule
\multirow{2}{*}[-0.8ex]{\parbox{2.4cm}{\centering \textbf{Models}}} &
\multirow{2}{*}[-0.8ex]{\parbox{1cm}{\centering \textbf{SBERT}}} &
\multirow{2}{*}[-0.8ex]{\parbox{1cm}{\centering \textbf{LLM-Judge}}} &
\multicolumn{4}{c}{\textbf{OBELS}} \\
\cmidrule{4-7}
 & & & \textbf{$E_{\text{func}}$} & \textbf{$E_{\text{dom}}$} & \textbf{$E_{\text{sem}}$} & \textbf{$T_{\text{ent}}$} \\ \midrule
Claude-opus-4-1 & \textbf{0.544} & \textbf{0.445} & 0.765 & \textbf{0.785} & \textbf{0.570} & \textbf{0.670} \\
Gemini-2.5-pro & 0.506 & 0.420 & 0.710 & 0.705 & 0.505 & 0.645 \\
GPT-5 & 0.470 & 0.240 & 0.710 & 0.715 & 0.535 & 0.670 \\
GPT-5-mini & 0.479 & 0.315 & 0.710 & 0.685 & 0.500 & 0.635 \\
GPT-4o & 0.492 & 0.415 & \textbf{0.770} & 0.735 & 0.520 & 0.640 \\  
GPT-4.1 nano  & 0.430 & 0.330 & 0.735 & 0.700 & 0.500 & 0.665 \\
Llama 3.1 8B  & 0.466 & 0.310 & 0.695 & 0.675 & 0.475 & 0.630 \\
Qwen 2.5-32B & 0.489 & 0.400 & 0.770 & 0.72 & 0.525 & 0.670 \\
Gemma 3-4B    & 0.429 & 0.280 & 0.675 & 0.670 & 0.475 & 0.635 \\
\bottomrule
\end{tabular}
\end{table}

\paragraphbe{Impact of Inference LLM.}
\autoref{tab:prompt_rec_models} compares inference engines for ICL-based prompt recovery, showing that model choice strongly affects reconstruction quality. Claude-opus-4-1 and Gemini-2.5-pro achieve the highest scores, while GPT-5, GPT-5-mini, and GPT-4o remain competitive. Larger models leak more information but incur higher cost and latency, especially in the Gemini, Claude, and GPT-5 families. Balancing accuracy, cost, and latency, we identify GPT-4o as the most practical option, offering strong recovery performance with lower overhead than top-scoring but slower systems.

\subsection{Trait Inference}
We assess how much and accurately an adversary can recover latent attributes of the user from a week-long browsing trace (see Section~\ref{sec:metrics}). 
The results quantify trait exposure using \textit{inference accuracy} (similarity score), where higher values mean more reliable adversarial inference.
Gemini 2.5 pro is used as the inference LLM.

\begin{figure}[t]
    \centering
    \includegraphics[width=\linewidth]{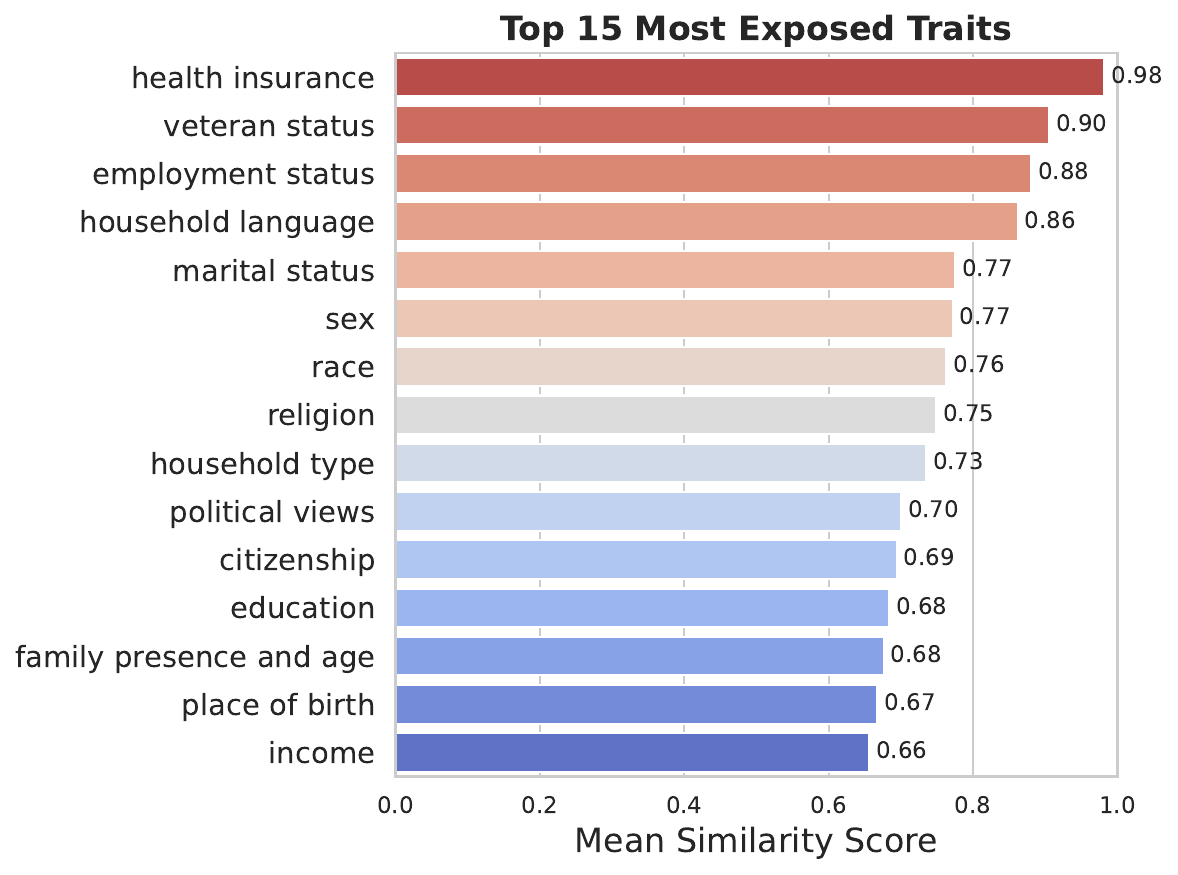}
    \captionsetup{skip=3pt}
    \caption{\textbf{Top 15 traits with highest exposure risk.} Inference risk is measured by mean similarity scores across all personas and sessions. Higher scores indicate stronger inference. }
    \label{fig:top15traits}
\end{figure}

\begin{figure*}[ht]
    \centering
    \includegraphics[width=\linewidth]{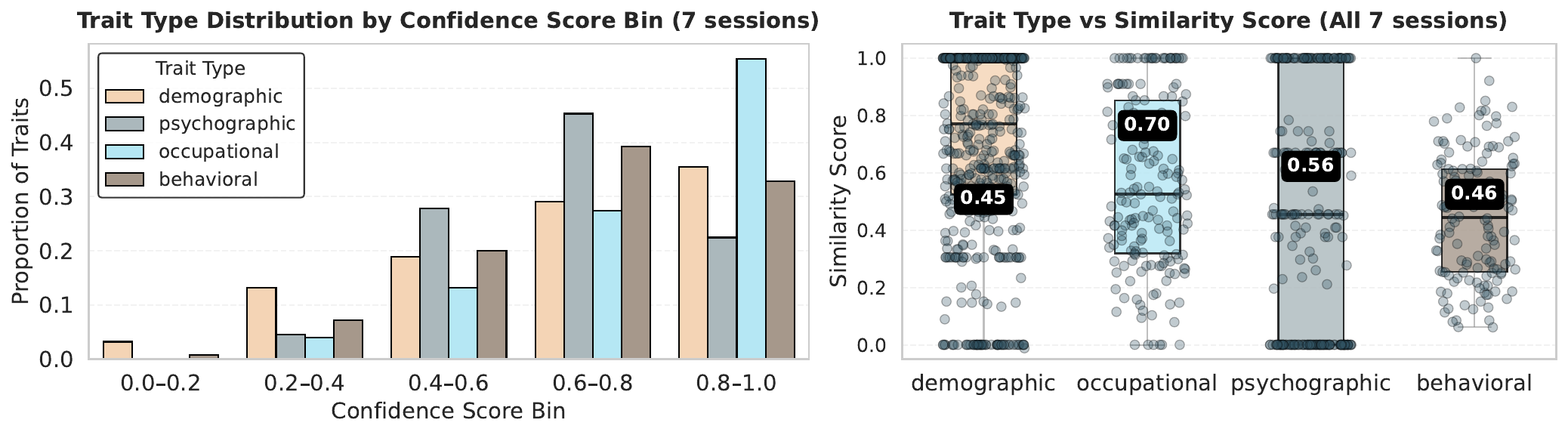}
    \captionsetup{skip=3pt}
    \caption{\textbf{Left:} Proportion of inferred traits in each confidence bin. Occupational and demographic traits are most often assigned high confidence (0.8–1.0), while the other two peak in middle ranges (0.6–0.8). \textbf{Right:} Similarity scores by trait category. Boxes show the interquartile range, black lines, squared numbers, and dots represent medians, means, and all data points, respectively.}
    \label{fig:confidence+inference}
\end{figure*}

\autoref{fig:top15traits} ranks the fifteen most exposed traits, showing that exposure varies by attribute. 
Health insurance (0.98), veteran status (0.90), employment status (0.88), and household language (0.86) are inferred with near-perfect accuracy, with employment status nearly as reliable as binary traits despite being multi-class. 
A second tier (marital status, sex, race, religion, household type) scores 0.73–0.77, while political views, citizenship, education, family presence/age, place of birth, and income range from 0.66–0.70, still well above random. 
Behavioral and psychographic traits do not appear in the top 15, consistent with their lower exposure in \autoref{fig:confidence+inference}. 
Overall, the results suggest that demographic and occupational traits are most vulnerable to leakage, while lifestyle and personality attributes remain less exposed but at nontrivial levels.

\begin{figure*}[ht]
    \centering
    \includegraphics[width=\linewidth]{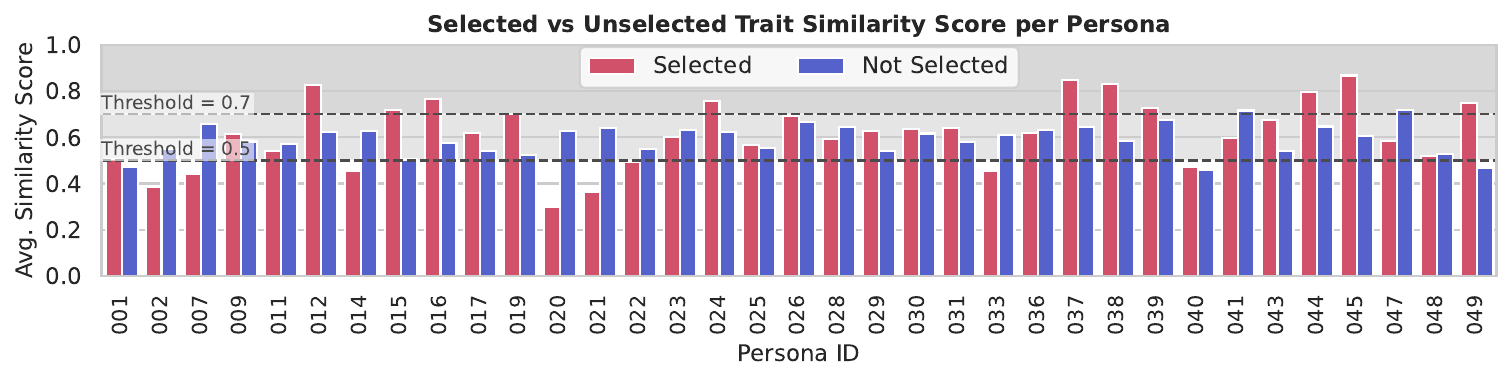}
    \captionsetup{skip=3pt}
    \caption{\textbf{Average similarity scores per persona for selected (red) and unselected (blue) traits.} Dashed lines at 0.5 and 0.7 indicate moderate and high-risk leakage thresholds. Many personas have selected traits exceeding 0.7, while unselected traits frequently remain above 0.5, reflecting moderate but non-negligible exposure.}
    \label{fig:selected-unselected-7sessions}
\end{figure*}

\autoref{fig:selected-unselected-7sessions} compares similarity scores for selected and unselected traits. 
Selected traits typically exceed 0.7, with some personas (012, 024, 037, 045) surpassing 0.8, indicating consistent high-risk leakage \cite{cann2025using}. 
Unselected traits generally fall below but often remain above 0.5, showing moderate exposure even when traits are not explicitly embedded. 
In a few cases (007, 025, 041), unselected traits approach or exceed selected ones, confirming leakage extends beyond explicitly mentioned attributes.

\autoref{fig:confidence+inference} compares model confidence and accuracy across categories. 
As shown in the left plot, occupational and demographic traits are assigned very high confidence (0.8--1.0), while the other two peak in the mid-confidence range (0.6--0.8).
The accuracy distributions (right plot) further distinguish the categories: occupational traits average 0.70 but with a lower median (0.53), psychographic traits are inferred with moderate accuracy (mean 0.56, median 0.46) but show extreme variability (spread 0--1).
Behavioral traits are uniformly moderate to weak (mean 0.46, median 0.45) with a narrowest spread of 0.25 to 0.61, and demographic traits show the highest median (0.90), but occasional errors lower the mean (0.44). 
Trait categories differ not only in accuracy but also in distribution: demographic traits are most reliably inferred, occupational traits are less consistent but often highly accurate, psychographic traits are highly variable, and behavioral traits are the least exposed.

\begin{figure}[ht]
    \centering
    \includegraphics[width=\linewidth]{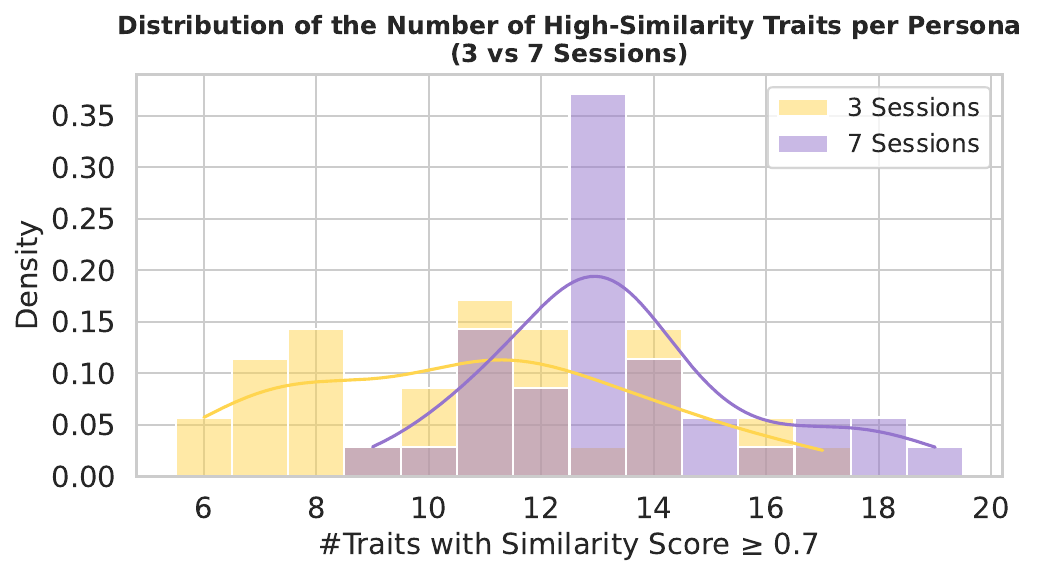}
    \captionsetup{skip=3pt}
    \caption{Distribution of traits per persona with similarity $\ge$ 0.7 under 3- and 7-session settings. The 7-session case shifts rightward, indicating more traits inferred at high similarity.}
    \label{fig:high-similarity-traits}
\end{figure}

\paragraphbe{Different Number of Sessions.}
As part of our ablation study, \autoref{fig:high-similarity-traits} shows the distribution of traits per persona with similarity scores above 0.7 under three vs.\ seven sessions. 
With three sessions (yellow), most personas reveal 6–10 traits with a long tail of higher counts. 
With seven sessions (purple), the distribution shifts rightward and concentrates around 12–14 traits, with some reaching 19. 
This demonstrates that additional sessions, and thus more domain traces, enable inference of more traits at high accuracy.

\autoref{tab:trait-category-summary} confirms consistent gains across categories, with psychographic (+16.7\%) and demographic (+15.4\%) improving most, followed by behavioral (+12.2\%) and occupational (+7.4\%) traits.  
Overall, longer observation windows strengthen inference, increasing both the number of high-similarity traits and average accuracy.

\begin{table}[t]
\centering
\renewcommand{\arraystretch}{0.9}
\caption{\textbf{Average similarity scores across trait categories.} $\Delta$ shows the improvement of full sessions over three sessions.}
\begin{tabular}{lccr}
\toprule
\textbf{Trait Category} & \textbf{3 Sessions} & \textbf{7 Sessions} & \textbf{$\Delta$ (\%)} \\
\midrule
Demographic    & 0.39 & 0.45 & +15.4\% \\
Occupational   & 0.68 & 0.73 & +7.4\% \\
Psychographic  & 0.48 & 0.56 & +16.7\% \\
Behavioral     & 0.41 & 0.46 & +12.2\% \\
\bottomrule
\end{tabular}
\label{tab:trait-category-summary}
\end{table}





Overall, our results show that metadata leakage enables both immediate and long-term risks. 
First, prompts can be reliably recovered regardless of whether traces are produced by proprietary or local LLM backbones, or by agents with varying exploration strategies. 
For trait inference, exposure levels differ across categories, but even traits not explicitly embedded in prompts can still be inferred at moderate to high accuracy, as shown by the recovery of unselected traits. 
Together, these findings demonstrate that domain traces alone reveal sensitive user intent and identity signals, underscoring metadata leakage as a significant privacy threat.

\section{Defense and Realistic Deployment} \label{sec:defense}

\paragraph{Motivation.} 
If a task requires visiting a unique domain, the defense should hide that interaction via VPNs or randomized background queries.
By contrast, if the information can be obtained without distinctive traces, blocking is preferable; for example, by querying an LLM directly or gathering information from Wikipedia.
We thus identify two conceptual mitigations for metadata-based inference attacks:
(1) \textbf{hiding traces}, where sensitive activity must occur but is camouflaged with plausible noise; and 
(2) \textbf{blocking traces}, where the activity is avoided by relying on alternative, less revealing sources. 

\subsection{Hiding via Decoy Prompts} \label{sec:defense-hiding}
Inspired by TrackMeNot~\cite{nissenbaum2009trackmenot}, for each real prompt the system generates and executes additional decoy prompts that remain within the same topical space but differ in framing, intent, or implied traits. 
Because OBELS evaluates whether reconstructions preserve intent, domain type, and entity, breaking this alignment is critical; otherwise, an adversary can still recover the user’s functional behavior despite wording changes. 
For instance, a prompt about ``recognizing depression and seeking medical help for a friend'' can be paired with a decoy like ``compare the effectiveness of yoga versus meditation for improving sleep quality'', which shifts both domain and entity, lowering OBELS score and obscuring the true goal. 
This ambiguity in observed traffic undermines an adversary’s ability to infer the user’s true objective, attributes, or target entity. 
We explore a lightweight approach using LLM-based decoy generation and a trait-aware variant that leverages virtual personas to introduce long-term behavioral divergence.

\paragraphbe{Mechanism.}
The defense operates in three stages. 
(1) \textbf{Decoy prompt generation}: an LLM produces multiple decoys that remain within the same topical space but differ in context, intent, or entity granularity, thereby obscuring both high-level intent and fine-grained attributes such as geographic focus or institutional target (e.g., a prompt on ``Swahili food'' may be paired with decoys on ``Italian cuisine'' or ``modern American dishes''). 
(2) \textbf{Trait-conflicting guidance} (optional): decoy generation can be steered by a virtual persona, with the system maintaining a rolling estimate of the user’s traits via lightweight keyword heuristics and selecting a persona that diverges on key dimensions (e.g., ideology or religion). 
These divergent decoys introduce consistent but misleading patterns, weakening long-term profiling. 
(3) \textbf{Concurrent execution}: real and decoy prompts are issued in parallel; while only the real output is returned to the user, all queries leave observable traces, blending plausible yet misleading activity into the traffic and complicating adversarial inference of the user’s true intent or identity.

\begin{table}[t]
\caption{\textbf{Effect of decoy prompts and domain shuffling on prompt leakage attacks.}}
\label{tab:prompt_rec_decoy_prompts_defense}
\centering
\small
\renewcommand{\arraystretch}{0.9}
\setlength\tabcolsep{1.8pt}
\begin{tabular}{>{\raggedright\arraybackslash}m{2.4cm}
                    >{\centering\arraybackslash}m{1cm}
                    >{\centering\arraybackslash}m{1cm}
                    *{4}{>{\centering\arraybackslash}m{.8cm}}}
\toprule
\multirow{2}{*}[-0.8ex]{\parbox{2.4cm}{\centering \textbf{Defense}}} &
\multirow{2}{*}[-0.8ex]{\parbox{1cm}{\centering \textbf{SBERT}}} &
\multirow{2}{*}[-0.8ex]{\parbox{1cm}{\centering \textbf{LLM-Judge}}} &
\multicolumn{4}{c}{\textbf{OBELS}} \\
\cmidrule{4-7}
 & & & \textbf{$E_{\text{func}}$} & \textbf{$E_{\text{dom}}$} & \textbf{$E_{\text{sem}}$} & \textbf{$T_{\text{ent}}$} \\ \midrule
Without Defense  & 0.492 & 0.415 & 0.770 & 0.735 & 0.520 & 0.640 \\ 
\cmidrule{1-7}
1 Decoy            & 0.390 & 0.190 & 0.610 & 0.615 & 0.420 & 0.545 \\ 
3 Decoys           & 0.376 & 0.210 & 0.645 & 0.615 & 0.420 & 0.555 \\ 
5 Decoys           & \textbf{0.368} & \textbf{0.175} & \textbf{0.565} & \textbf{0.590} & \textbf{0.395} & \textbf{0.500} \\ 
1 Decoys + shuffle & 0.455 & 0.285 & 0.725 & 0.715 & 0.510 & 0.610 \\ 
3 Decoys + shuffle & 0.415 & 0.235 & 0.685 & 0.645 & 0.455 & 0.610 \\ 
5 Decoys + shuffle & 0.403 & 0.220 & 0.650 & 0.695 & 0.465 & 0.570 \\ 
\bottomrule
\end{tabular}
\end{table}

\paragraphbe{Impact of Defense on Prompt Leakage Attack.}
\autoref{tab:prompt_rec_decoy_prompts_defense} summarizes the prompt recovery performance under defense applied. The defense injects a varying number of decoy prompts, each contributing its own set of visited domains, so that the agent runs for both the original and decoy prompts simultaneously. Original domains are randomly interleaved within these sets while preserving their internal order, as shown in rows 2–4 of the table. To model a stronger defender, we also consider a condition where all domains, both original and decoy, are fully shuffled (rows 5--7).

The results show a consistent reduction in inference accuracy once decoys are added. With only one decoy prompt, SBERT drops by 0.10 and LLM-Judge by 0.23 relative to the baseline. More decoys generally strengthen the defense, with the largest effect observed at five decoys. When domains are shuffled, attack performance recovers somewhat for semantic and domain-level metrics, but still remains below baseline.

\begin{table}[t]
\centering
\renewcommand{\arraystretch}{0.9}
\caption{\textbf{Median inference accuracy across trait categories.}}
\begin{tabular}{lccr}
\toprule
\textbf{Trait Category} & \textbf{No Defense} & \textbf{w. Defense} & \textbf{$\Delta$ (\%)} \\
\midrule
Demographic    & 0.4440 & 0.4080 & $-8.1\%$ \\
Occupational   & 0.8975 & 0.6790 & $-24.3\%$ \\
Psychographic  & 0.5250 & 0.5030 & $-4.2\%$\\
Behavioral     & 0.4550 & 0.4550 & $0.0\%$ \\
\bottomrule
\end{tabular}
\label{tab:defense-trait-inference}
\end{table}

\paragraphbe{Impact of Defense on Trait Inference.}
We evaluate the defense by re-running the trait inference attack with decoy prompts injected. 
In this setting, each real prompt triggers a concurrent decoy prompt from a conflicting virtual persona, selected based on the defender-side trait estimate.
\autoref{tab:defense-trait-inference} reports median similarity scores between inferred and ground-truth traits across four categories. 
The defense yields the largest median reduction for occupational traits ($-24.3\%$), followed by demographic traits ($-8.1\%$). 
Psychographic traits drop slightly ($-4.2\%$), while behavioral traits remain unchanged.
Additional results are provided in \autoref{app:additional-defense}.

\subsection{Blocking via Alternative Sources} \label{sec:defense_blocking}
A complementary defense is to prevent sensitive traces from arising in the first place. 
When a user’s objective can be met without contacting uniquely identifying domains, the agent can instead draw from large multipurpose repositories (e.g., Wikipedia, StackExchange, Reddit) or from the LLM’s internal knowledge. 
By retrieving information from resources that serve diverse intents or by leveraging the model’s internal knowledge, the resulting traffic no longer maps cleanly to a specific user goal. 

From the adversary’s perspective, this strategy effectively reduces visibility by collapsing domain diversity, yielding traces similar to those observed by weaker observers that see limited or non-discriminative domain information. 
This ``blocking'' approach therefore both mitigates leakage and illustrates how reduced observability weakens inference, reducing the adversary’s ability to associate domain visits with private attributes.


\begin{table}[t]
\caption{\textbf{Prompt recovery under different trace visibility.} URL-level traces leak the most; adding timing metadata shows no consistent benefit. }
\label{tab:prompt_rec_matadata}
\centering
\small
\renewcommand{\arraystretch}{0.9}
\setlength\tabcolsep{.8pt}
\begin{tabular}{>{\raggedright\arraybackslash}m{2.9cm}
                    >{\centering\arraybackslash}m{1cm}
                    >{\centering\arraybackslash}m{1cm}
                    *{4}{>{\centering\arraybackslash}m{.8cm}}}
\toprule
\multirow{2}{*}[-0.8ex]{\parbox{2.4cm}{\centering \textbf{Trace Visibility}}} &
\multirow{2}{*}[-0.8ex]{\parbox{1cm}{\centering \textbf{SBERT}}} &
\multirow{2}{*}[-0.8ex]{\parbox{1cm}{\centering \textbf{LLM-Judge}}} &
\multicolumn{4}{c}{\textbf{OBELS}} \\
\cmidrule{4-7}
 & & & \textbf{$E_{\text{func}}$} & \textbf{$E_{\text{dom}}$} & \textbf{$E_{\text{sem}}$} & \textbf{$T_{\text{ent}}$} \\ \midrule
Domains & 0.492 & 0.415 & 0.770 & 0.735 & 0.520 & 0.640 \\ 
Domains + Timing   & 0.486 & 0.455 & 0.725 & 0.695 & 0.525 & 0.640 \\
URLs (100\%)     & \textbf{0.624} & \textbf{0.600} & 0.810 & \textbf{0.805} & 0.620 & \textbf{0.695} \\
Partial URLs (80\%)    & 0.618 & 0.590 & 0.795 & 0.795 & \textbf{0.635} & 0.680 \\
Partial URLs (60\%)    & 0.622 & 0.590 & \textbf{0.820} & 0.755 & 0.585 & 0.690 \\
\bottomrule
\end{tabular}
\end{table}

\begin{table}[t]
\caption{\textbf{Leakage vs. utility under varying visibility. }
Reducing visibility lowers prompt recovery accuracy but leaves the report quality nearly unchanged.}
\label{tab:prompt_rec_utility}
\centering
\renewcommand{\arraystretch}{0.9}
\small
\setlength\tabcolsep{2pt}
\begin{tabular}{>{\raggedright\arraybackslash}m{1.1cm}
                    >{\centering\arraybackslash}m{1cm}
                    >{\centering\arraybackslash}m{1cm}
                    >{\centering\arraybackslash}m{1cm}
                    *{4}{>{\centering\arraybackslash}m{.8cm}}}
\toprule
\multirow{2}{*}[-0.8ex]{\parbox{1.1cm}{\centering \textbf{Visibility}}} &
\multirow{2}{*}[-0.8ex]{\parbox{1.0cm}{\centering \textbf{Utility}}} &
\multirow{2}{*}[-0.8ex]{\parbox{1cm}{\centering \textbf{SBERT}}} &
\multirow{2}{*}[-0.8ex]{\parbox{1cm}{\centering \textbf{LLM-Judge}}} &
\multicolumn{4}{c}{\textbf{OBELS}} \\
\cmidrule{5-8}
 & & & & \textbf{$E_{\text{func}}$} & \textbf{$E_{\text{dom}}$} & \textbf{$E_{\text{sem}}$} & \textbf{$T_{\text{ent}}$} \\ \midrule
100\%  & 8.55 & 0.492 & 0.415 & 0.770 & 0.735 & 0.520 & 0.640 \\ 
80\%   & 8.72& 0.491 & 0.415 & 0.750 & 0.725 & 0.515 & 0.645 \\
60\%   & 8.54& 0.489 & 0.355 & 0.765 & 0.730 & 0.520 & 0.650\\
40\%  & 8.50 & 0.453 & 0.350 & 0.765 & 0.700 & 0.515 & 0.650\\
20\%  & 8.60& 0.438 & 0.300 & 0.725 & 0.695 & 0.475 & 0.615\\
10\%  & 8.54 & 0.375 & 0.225 & 0.610 & 0.555 & 0.395 & 0.485\\
5\% & 8.32 & 0.319 & 0.185 & 0.530 & 0.455 & 0.335 & 0.395 \\
2\% & 7.65 & 0.238 & 0.095 & 0.345 & 0.410 & 0.275 & 0.315\\
\bottomrule
\end{tabular}
\end{table}

\autoref{tab:prompt_rec_matadata} and \autoref{tab:prompt_rec_utility} together show that visibility of browsing traces strongly shapes leakage but has little effect on utility. Utility is how much the WRA’s output helps a human reader reach their research goals. We measure it with an LLM using the prompt template. 
URL-level traces leak substantially more than domain-only traces, making prompt inference easier and more accurate. Reducing visibility (e.g., observing only 80\% or 60\% of domains) lowers leakage but still exposes useful information. Adding timing metadata to domains yields no consistent gains, suggesting that domain identity is more informative than temporal order. In contrast, the utility of the generated reports remains relatively stable across visibility levels down to 10\%. However, beyond this threshold, utility begins to drop, revealing diminishing returns from additional domain coverage. This suggests that \emph{many extra domains contribute little to report quality while disproportionately increasing exposure to adversaries}. Thus, WRA behaviors that aggressively visit many domains are especially problematic: they create substantial leakage risks while providing only marginal gains in utility.



\section{Discussion and Conclusion}
\paragraphbe{Privacy and Security Implications.}
Our results show that metadata leakage is an inherent risk for locally deployed web and research agents. We find that design choices---including domain exploration breadth, backbone model capacity, and orchestration logic---directly affect the amount of information exposed. 
Substituting locally deployable models does not eliminate leakage: domain-level traffic is structurally unavoidable. 
This shifts the attack vector from \textit{content} to \textit{behavior}. 
Whereas prompt injection and data exfiltration could be mitigated through isolation~\cite{tsai2025contextual}, network-level traces cannot be trivially suppressed. 
At scale, such leakage enables profiling by advertisers, ISPs, or state actors, raising unique questions about how to balance agent utility with user privacy. 

\paragraphbe{Defense Effectiveness and Limitations.}
Our study has several limitations. 
We rely on proxy datasets and synthetic personas with limited session counts; longer-term aggregation or multimodal attacks may amplify risks. 
Defense evaluation is confined to decoys and blocking, and cost/latency trade-offs are only partially measured. 
Nonetheless, our results show that decoy prompts and trait-conflicting personas reduce both prompt recovery and trait inference accuracy, while blocking strategies help when tasks can rely on broad, non-unique sources (e.g., Wikipedia) without sacrificing utility. 
Yet defenses remain partial: adversaries still recover useful signals, and decoys introduce overhead in latency, bandwidth, and behavioral realism. 
As with prior obfuscation systems, mitigation is meaningful but incomplete; full network-level protections (e.g., VPNs or anonymity systems for agents~\cite{pham2024proxygpt})  remain the only robust safeguard.

\paragraphbe{Future Directions and Open Challenges.}
Future work should explore hybrid defenses that combine obfuscation, blocking, and timing perturbations; study long-term aggregation attacks across extended user histories; and integrate formal privacy guarantees such as differential privacy or traffic-analysis resistance directly into agent design. 
Multi-modal agents introduce additional risks, for example, as passed multi-modal content could reveal even more information~\cite{schuster2017beauty}. 
Overall, progress requires viewing AI agents and the web as a coupled ecosystem rather than orthogonal components.

\paragraphbe{Conclusion.}
We show prompt and trait leakage from network traces of local WRAs. 
Even with defenses, residual exposure persists because traces are structurally generated and behaviorally rich. 
Our study demonstrates both the feasibility of inference attacks and the incompleteness of current defenses, underscoring the need for systemic, privacy-aware agent design. 
Until such frameworks are developed, strong network-level protections remain the only reliable safeguard.

\section*{Acknowledgments}
The work was partially supported by Schmidt Sciences SAFE-AI program and by the NSF grants 2333965 and 2131910. 

\appendix
\section{Ethical Considerations}
Our research examines how WRAs can inadvertently leak sensitive information through metadata such as domain access patterns, timing, and interaction traces. 
While our goal is to strengthen privacy protections for agent-based systems, we recognize the potential for harmful use if such inference techniques were deployed maliciously.

\paragraphb{Stakeholders.} 
We identify four stakeholder groups affected by this research: 
(1) End-users of WRAs, who may have personal traits, goals, or intentions inferred without their awareness. 
This risk is not uniform; individuals in politically sensitive contexts, marginalized communities, or high-surveillance environments may face disproportionately higher harm. 
(2) Developers and framework maintainers, who may need to modify or harden system architectures, logging defaults, and data retention practices in response to identified risks. 
(3) Organizations deploying agents (e.g., educational platforms, customer service systems), which may unknowingly violate privacy expectations or regulatory obligations if leakage is not addressed. 
(4) The broader public, which has a collective interest in ensuring that emerging agent ecosystems evolve in privacy-preserving and autonomy-respecting directions.

\paragraphb{Mitigating Potential Harm.} 
To reduce risks to end-users, we designed our study so that no real user data were collected or analyzed; all experiments use public benchmark datasets (FEDWEB13, SESSION14, DD16) and synthetic personas, ensuring that no identifiable individuals can be profiled or re-identified. 
To support developers and framework maintainers, we present concrete mitigation strategies, such as trace obfuscation and domain suppression, that can be incorporated into WRA architectures to reduce metadata leakage. 
For organizations that deploy agents in educational, corporate, or customer-service environments, we do not release any tools for live traffic monitoring, and the accompanying code is limited to offline, dataset-based replay experiments, preventing misuse in operational settings. 
Finally, by identifying these risks early and providing actionable defenses, the work serves broader public interests by promoting privacy-preserving WRA design, rather than end-user surveillance or behavioral profiling as agents become widely deployed.

\paragraphb{Researcher Well-being and Responsibilities.} 
Our research did not involve direct interaction with online communities, covert participation, or exposure to disturbing or sensitive content. However, we continuously discussed the dual-use concerns throughout the project and made design choices aimed at minimizing the potential for misuse, for example, limiting our evaluation to public datasets and not designing tools that can be directly applied to live user traffic.

\paragraphb{Societal Impact and Justification.} 
As WRAs become integrated into productivity tools, classrooms, workplaces, and personal browsing workflows, metadata-based inference risks could scale rapidly and invisibly. By identifying and characterizing this attack vector under controlled conditions, we enable developers, regulators, and practitioners to recognize and address these vulnerabilities early, before they are exploited in real-world deployments. By focusing on public data, transparent methodology, and concrete mitigation guidance, we ensure that the research advances privacy-preserving agent design while minimizing the likelihood of harmful use.

\section{Open Science}
Implementations, evaluation code, prompts, and the synthetic datasets derived from our study will be made publicly available. 
Code and prompts available at: \url{https://github.com/umass-aisec/wra}.

\bibliographystyle{plain}
\bibliography{references}

\section{Experimental Setup Details}
\subsection{Dataset Details} \label{app:dataset_details}
\paragraphbe{Prompt Recovery.}
For all TREC datasets used in the prompt recovery task, we create \textbf{-DR} variants by inputting each original prompt with \texttt{suggested\_rewriting\_prompt}\cite{alwell2025introduction}, to GPT-4.1.  
The rewritten prompts are used for the Deep Research API runs to match its prompt handling, and all evaluations are scored against the original, unmodified prompts.

\textbf{TREC FedWeb 2013 Topics} contains 50 topics originally designed to evaluate federated web search. We concatenate the \texttt{<description>} and \texttt{<narrative>} fields to form a single prompt, and rewritten with GPT-4.1 (FEDWEB13-DR).

\textbf{TREC Session 2014 Topics} contains 60 realistic user topic descriptions spanning domains such as travel, health, and education. 
We use the original \texttt{description} fields as user prompts and their rewritten counterparts (SESSION14-DR).

\textbf{TREC Dynamic Domain 2016 Truth Data} provides 53 topics focused on interactive information retrieval in evolving domains such as medicine and cybersecurity. 
We concatenate the \texttt{<description>} and \texttt{<narrative>} fields and rewrite them with GPT-4.1 (DD16-DR).

For each dataset, 20 prompts are reserved for evaluation, with the remainder used for training and ICL examples. \\

\paragraphbe{Trait Inference Personas.}
We use \textbf{SynthLabsAI/PERSONA\_subset} to simulate benign users who use a web agent over time.
The dataset includes 997 synthetic user profiles annotated with 32 traits in four categories (demographic, occupational, psychographic, and behavioral). 
We adopt a type-aware scoring for each trait for evaluation: 
\begin{compactitem}
    \item \textit{Numeric traits} (e.g., age, income) are evaluated using a normalized absolute difference: smaller relative errors yield higher scores, linearly decreasing toward zero. Age differences are scaled by 30, and income by 200K.
    \item \textit{Ordinal traits} (e.g., big five scores) are mapped to integer levels and scored by normalized distance on a 5-point scale. Closer ratings yield higher similarity scores.
    \item \textit{Categorical traits} (e.g., gender, race) use exact match for single-token values; multi-word categories are scored using SBERT score to account for semantic equivalence.
    \item \textit{Free-text traits} (e.g., job description, personal time) are assessed using the SBERT score, which computes semantic overlap via contextual token embeddings.
\end{compactitem}

In our experiments, we sample 50 personas. 
For each, including those reserved for ICL examples, we randomly select 5 traits and use GPT-4o ($T=0.7$) to generate 21–35 trait-revealing prompts that explicitly or implicitly embed the traits.
The produced traces serve as the input for trait inference.

\begin{table}[t]
\centering
\small
\captionsetup{skip=6pt}
\setlength\tabcolsep{1.2pt}
\renewcommand{\arraystretch}{0.9}
\caption{Few-shot ICL configuration for each attack task.}
\label{tab:icl-config}
\begin{tabular}{>{\raggedright\arraybackslash}m{2.18cm}
                    >{\centering\arraybackslash}m{.8cm}
                    >{\centering\arraybackslash}m{2.9cm}
                    >{\centering\arraybackslash}m{2.3cm}}
\toprule
\textbf{Task} & \textbf{ICL Shots} & \textbf{Output Format} & \textbf{Prompt Details} \\
\midrule
Prompt Recovery & 5 & Natural language query & Trace $\rightarrow$ prompt \\
Trait Inference & 3 & Structured list        & Trace $\rightarrow$ trait list \\
\bottomrule
\end{tabular}
\end{table}

\subsection{Agent and LLM Backbone Choice} \label{app:trace_collection}
When browsing is executed locally, encrypted outbound traffic remains visible to an on-path adversary; when browsing is server-side (e.g., OpenAI’s Operator), only connections to the provider’s infrastructure are exposed. 
We collect traces using both open-source and proprietary WRAs.

\textbf{GPT Researcher \cite{elovic2023gptresearcher}.}
GPT Researcher is a research agent designed for multi-hop investigation that decomposes prompts into sub-questions, gathers evidence, and synthesizes reports. 
We run it in \texttt{report\_type=``deep''} mode, enabling multi-turn exploration across sources. 
GPT Researcher employs different LLMs for sub-tasks: \texttt{FAST\_LLM} for lightweight operations (e.g., GPT-4o-mini, DeepSeek-V3-0324), \texttt{SMART\_LLM} for reasoning and report generation (e.g., GPT-4.1, Mistral-7B-Instruct v0.3), and \texttt{STRATEGIC\_LLM} for higher-level planning (e.g., o4-mini, Mistral-Nemo-12B). 
While the default configuration uses GPT-4, we also evaluate local LLM combinations to assess leakage outside proprietary ecosystems.

\textbf{Browser-Use \cite{browser_use2024}.}
Browser-Use is an open-source agent that automates browsing through a local browser. 
By default, it issues Google queries via \texttt{search\_web}, which frequently triggers reCAPTCHA; we therefore configure it to use Bing with automatic fallback. 
We further modify its system prompt to require at least five distinct page visits before summarization, preventing it from shortcutting tasks with prior knowledge or superficial searches. 
After each run, we extract domains and timestamps from the browser’s recorded history to construct ordered traces. 
This setup makes Browser-Use suitable for evaluating how web agents' lightweight browsing patterns leak information.

\textbf{AutoGen \cite{wu2023autogen}.}  
We configure AutoGen as a two-agent system consisting of an \texttt{AssistantAgent} and a \texttt{MultimodalWebSurfer}. 
Compared to Browser-Use, AutoGen is faster and more controllable, making it useful for both tasks.
By default, AutoGen tends to summarize results directly without clicking through, so we enforce the same rule, visiting at least five pages. 
To collect metadata, we extend the web surfer into a custom \texttt{LoggingWebSurfer} that hooks Playwright APIs to record domains, timestamps, and IPs for all network events. 
Unlike GPT Researcher or Browser-Use, AutoGen logs not only primary domains but also auxiliary requests (ads, trackers, analytics), producing noisier but realistic traces that an actual network observer would see. 
We run AutoGen with GPT-4o for prompt recovery and Gemini 2.0 Flash for trait inference.

\textbf{OpenAI Deep Research API.}
As a proprietary baseline, we evaluate the OpenAI Deep Research API. 
Unlike the interactive ChatGPT Deep Research interface, the API is stateless and relies entirely on instruction-rich inputs. 
To align its behavior with the UI, we provide \textbf{-DR} prompt variants (e.g., SESSION14-DR), rewritten to simulate the iterative refinement that would have been normally triggered by user interaction.  
Because browsing occurs remotely, only provider-level connections are visible locally; we nonetheless analyze these traces to benchmark inference difficulty against a high-capability proprietary system.

\paragraphbe{Rationale for Agent and Model Selection.}
We evaluate both research-oriented (GPT Researcher) and web-oriented (Browser-Use, AutoGen) agents to capture variation in browsing behavior and leakage. 
Each open-source agent is paired with both proprietary (GPT-4o, Gemini 2.0 Flash) and open-source (DeepSeek-V3, Mistral family) LLMs, reflecting realistic local deployments where users may expect stronger privacy guarantees. 
The Deep Research API serves as a commercial baseline.
We deliberately exclude larger open-source models (e.g., LLaMA 3.1 70B, DeepSeek-R1), as they require substantial computational resources and often bypass web search by relying on internal knowledge, making them less suitable for our browsing-driven evaluation. 
Together, these choices ensure coverage of attacker-visible metadata across diverse agents and realistic local deployment settings.

\subsection{Inference Configuration} \label{app:inference_config}
We evaluate our inference attacks using LLMs as an inference engine for few-shot ICL. 
A summary of ICL configurations is provided in \autoref{tab:icl-config}.

\paragraphbe{Prompt Recovery.}
Each ICL prompt consists of five (trace, prompt) pairs, where the trace includes an ordered domain sequence and associated timing information, followed by an unseen trace for prediction. 
The format is fully natural language, with traces presented as newline-separated domain lists (and timing when present). 
Examples are drawn from distinct topics to prevent overfitting and to test generalization across different intent types. 
This format encourages the model to infer high-level task semantics from the structure, content, and temporal progression of visited domains.

\paragraphbe{Trait Inference.}
For trait inference, we adopt a 3-shot ICL format. Each example provides a domain trace and corresponding persona traits.
The reduced shot count reflects the greater regularity of trait-disclosing behavior over time and helps minimize prompt length for complex structured outputs. 
The model is asked to generate a structured trait list in the format \texttt{- Trait: Value}, and is explicitly instructed to infer as many traits as possible, even under partial evidence. 

\begin{table}[t]
\caption{\textbf{Prompt recovery: 5-shot vs. contrastive ICL.} }
\label{tab:prompt_rec_main_method}
\centering
\small
\renewcommand{\arraystretch}{0.85}
\setlength\tabcolsep{1.5pt}
\begin{tabular}{>{\raggedright\arraybackslash}m{2.1cm}
                    >{\centering\arraybackslash}m{1cm}
                    >{\centering\arraybackslash}m{1.4cm}
                    *{4}{>{\centering\arraybackslash}m{.8cm}}}
\toprule
\multirow{2}{*}[-0.8ex]{\parbox{2.1cm}{\centering \textbf{Type of ICL}}} &
\multirow{2}{*}[-0.8ex]{\parbox{1cm}{\centering \textbf{SBERT}}} &
\multirow{2}{*}[-0.8ex]{\parbox{1.4cm}{\centering \textbf{LLM-Judge}}} &
\multicolumn{4}{c}{\textbf{OBELS}} \\
\cmidrule{4-7}
 & & & \textbf{$E_{\text{func}}$} & \textbf{$E_{\text{dom}}$} & \textbf{$E_{\text{sem}}$} & \textbf{$T_{\text{ent}}$} \\ \midrule
5-shot ICL   & 0.492 & \textbf{0.415} & \textbf{0.770} & 0.735 & 0.520 & 0.640 \\
Simple (1 neg)   & \textbf{0.495} & 0.405 & 0.740 & 0.745 & \textbf{0.525} & \textbf{0.645} \\
Simple (3 neg)   & 0.490 & 0.355 & 0.755 & \textbf{0.750} & 0.510 & 0.625\\
QF (1 neg)   & 0.493 & 0.385 & 0.765 & 0.680 & 0.520 & 0.590 \\
QF (3 neg)  & 0.492 & 0.355 & 0.750 & 0.740 & 0.505 & 0.640 \\
\bottomrule
\end{tabular}
\end{table}

\begin{table}[t]
\caption{\textbf{Prompt recovery: example selection strategies.} }
\small
\renewcommand{\arraystretch}{0.85}
\label{tab:prompt_rec_sel}
\centering
\setlength\tabcolsep{1pt}
\begin{tabular}{>{\raggedright\arraybackslash}m{2.6cm}
                    >{\centering\arraybackslash}m{1cm}
                    >{\centering\arraybackslash}m{1cm}
                    *{4}{>{\centering\arraybackslash}m{.8cm}}}
\toprule
\multirow{2}{*}[-0.8ex]{\parbox{2.6cm}{\centering \textbf{Ex Selection Strategy}}} &
\multirow{2}{*}[-0.8ex]{\parbox{1cm}{\centering \textbf{SBERT}}} &
\multirow{2}{*}[-0.8ex]{\parbox{1cm}{\centering \textbf{LLM-Judge}}} &
\multicolumn{4}{c}{\textbf{OBELS}} \\
\cmidrule{4-7}
 & & & \textbf{$E_{\text{func}}$} & \textbf{$E_{\text{dom}}$} & \textbf{$E_{\text{sem}}$} & \textbf{$T_{\text{ent}}$} \\ \midrule

Random   & \textbf{0.492} & \textbf{0.415} & \textbf{0.770} & 0.735 & \textbf{0.520} & \textbf{0.640} \\
Embedding-based    & 0.481 & 0.375 & 0.740 & \textbf{0.755} & 0.515 & 0.605 \\
\bottomrule
\end{tabular}
\end{table}

\begin{table}[t]
\caption{\textbf{Prompt recovery: different example orderings} }
\label{tab:prompt_rec_ordering}
\small
\renewcommand{\arraystretch}{0.85}
\centering
\setlength\tabcolsep{3pt}
\begin{tabular}{>{\raggedright\arraybackslash}m{1.8cm}
                    >{\centering\arraybackslash}m{1cm}
                    >{\centering\arraybackslash}m{1cm}
                    *{4}{>{\centering\arraybackslash}m{.8cm}}}
\toprule
\multirow{2}{*}[-0.8ex]{\parbox{1.8cm}{\centering \textbf{Ex Ordering}}} &
\multirow{2}{*}[-0.8ex]{\parbox{1cm}{\centering \textbf{SBERT}}} &
\multirow{2}{*}[-0.8ex]{\parbox{1cm}{\centering \textbf{LLM-Judge}}} &
\multicolumn{4}{c}{\textbf{OBELS}} \\
\cmidrule{4-7}
 & & & \textbf{$E_{\text{func}}$} & \textbf{$E_{\text{dom}}$} & \textbf{$E_{\text{sem}}$} & \textbf{$T_{\text{ent}}$} \\ \midrule
Random      & 0.481 & 0.375 & 0.740 & \textbf{0.755} & 0.515 & 0.605 \\
Ascending   & \textbf{0.493} & \textbf{0.415} & 0.725 & 0.735 & \textbf{0.525} & \textbf{0.645} \\
Descending   & 0.481 & 0.380  & \textbf{0.755} & 0.745 & 0.495 & 0.620 \\
\bottomrule
\end{tabular}
\end{table}

\begin{figure*}[t]
    \centering
    \includegraphics[width=.93\linewidth]{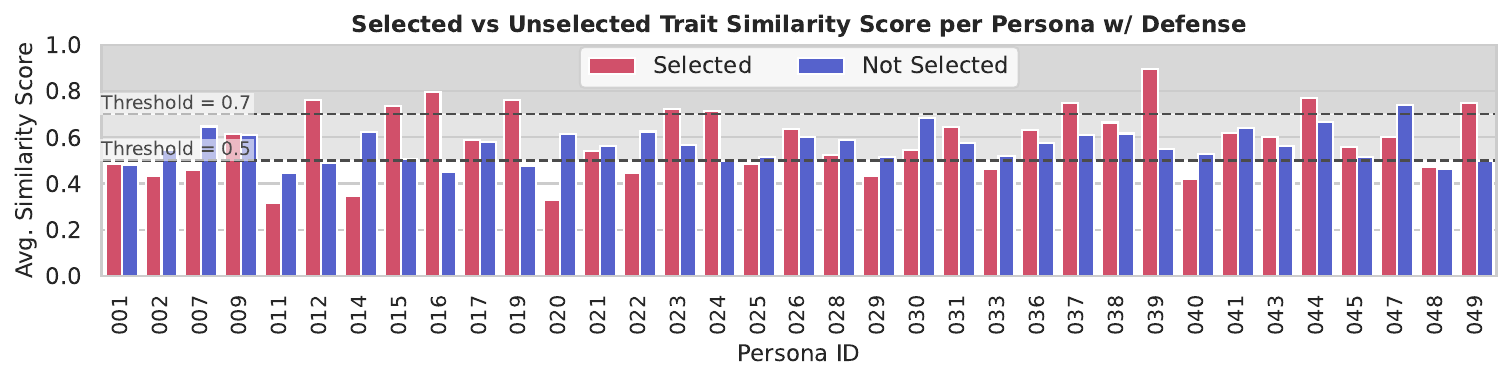}
    \captionsetup{skip=4pt}
    \caption{Average inference accuracy for selected vs. unselected traits per persona, showing generally lower accuracy than \autoref{fig:selected-unselected-7sessions} with the proposed virtual persona defense applied.}
    \label{fig:defense_per_persona}
\end{figure*}

\begin{figure*}[t]
    \centering
    \includegraphics[width=.93\linewidth]{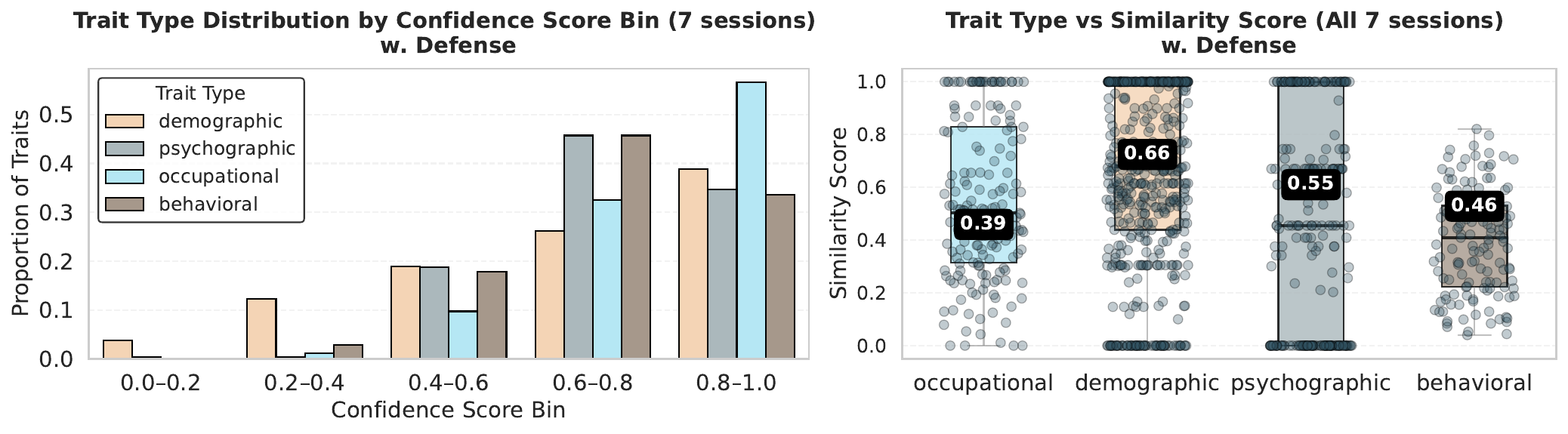}
    \captionsetup{skip=4pt}
    \caption{Trait-level breakdown of inference performance with the virtual persona defense.
    Confidence distributions remain similar to \autoref{fig:confidence+inference}, but average similarity scores are lowered across all categories.}
    \label{fig:defense_confidence+inference}
\end{figure*}

\section{Additional Prompt Recovery Results} \label{app:additional-prompt_recon}

\paragraphbe{Effect of ICL Configurations.}
Contrastive ICL\cite{gao2024customizing} showed that adding negative examples to the in-context setup can improve performance, even when negatives must be synthesized by the LLM. 
In our baseline, each 5-shot prompt includes five (trace, prompt) pairs followed by a new trace for prediction.
Contrastive variants augment these with one or three negatives, either in a simple setup or in our proposed \textbf{Quality-Filtered (QF) contrastive ICL}, where negatives are regenerated until their SBERT similarity to the original prompt falls below a threshold, ensuring sufficient dissimilarity.  

As shown in \autoref{tab:prompt_rec_main_method}, adding negatives did not improve overall results. While domain-type alignment ($E_{\text{dom}}$) occasionally improved, most other metrics decreased. Even with larger numbers of negatives and QF filtering, performance remained below the 5-shot baseline. We attribute this to the difficulty of the prompt inference task: the model may already operate near its limits, so additional negatives increase confusion rather than sharpening topical discrimination.

\paragraphbe{Effect of Example Selection.}
We compare random versus embedding-based selection of ICL examples. In the random strategy, examples are sampled from the training set for each test instance. For embedding-based selection, all training and test traces are encoded with the \texttt{all-MiniLM-L6-v2}\footnote{Fine-tuned on 1B sentence pairs with a contrastive objective, this model maps each trace into a 384-dimensional vector space.} sentence transformer, and the training examples with the highest cosine similarity to the test trace are chosen. As shown in \autoref{tab:prompt_rec_sel}, embedding-based selection does not improve prompt recovery accuracy and, in some metrics, performs slightly worse, suggesting that random sampling, by providing more diverse examples, is the more effective strategy.

\paragraphbe{Effect of Example Ordering.}
We test whether the ordering of embedding-selected examples influences ICL performance, comparing random order, ascending similarity, and descending similarity. As shown in \autoref{tab:prompt_rec_ordering}, ordering has little effect: scores remain close across all metrics. This suggests that once relevant examples are chosen, their sequence contributes minimally to recovery quality.

\section{Additional Virtual Persona Defense Results} \label{app:additional-defense}

\autoref{fig:defense_per_persona} and \autoref{fig:defense_confidence+inference} provide additional experimental results of the proposed virtual persona defense beyond the median similarity score reductions reported in the main text.

\end{document}